\documentclass[aps,prl,twocolumn,preprintnumbers,amsmath,amssymb,superscriptaddress]{revtex4-2}
\usepackage{amsmath,graphicx}
\usepackage[utf8]{inputenc}
\usepackage[T1]{fontenc}
\usepackage{xcolor}
\usepackage{textcomp}
\usepackage{bm}

\usepackage{siunitx}
\usepackage{physics}
\usepackage{amsmath}
\usepackage{tikz}
\usepackage{mathdots}
\usepackage{yhmath}
\usepackage{cancel}
\usepackage{color}
\usepackage{array}
\usepackage{multirow}
\usepackage{amssymb}
\usepackage{gensymb}
\usepackage{tabularx}
\usepackage{extarrows}
\usepackage{booktabs}
\usetikzlibrary{fadings}
\usetikzlibrary{patterns}
\usetikzlibrary{shadows.blur}
\usetikzlibrary{shapes}

\usepackage{color}
\definecolor{LinkColor}{rgb}{0.75,0.0,0.2}

\usepackage{hyperref}
\hypersetup{
	pdfauthor={good guys},
	pdftitle={good title},
	colorlinks=true,
	citecolor=LinkColor,
	linkcolor=LinkColor,
	urlcolor=LinkColor,
}

\usepackage{listings}
\definecolor{lightgray}{gray}{1}

\usepackage{theorem}

\usepackage{tikz}

\newcommand{\nc}{\newcommand}
\nc{\braoprket}[3]{\langle#1|#2|#3\rangle}
\nc{\opn}[1]{\operatorname{#1}}
\nc{\avg}[1]{\langle#1\rangle}
\nc{\ketbrasame}[1]{|#1\rangle\!\langle#1|}
\nc{\swap}{\opn{SWAP}}
\nc{\E}{\mathbb{E}}
\nc{\Var}{\opn{Var}}
\nc{\dg}{\dagger}

\usepackage[normalem]{ulem}
\begin{document}
\title{Gapless Symmetry-Protected Topological States in Measurement-Only Circuits}

\author{Xue-Jia Yu}
\altaffiliation{The first two authors contributed equally.}
\affiliation{Department of Physics, Fuzhou University, Fuzhou 350116, Fujian, China}
\affiliation{Fujian Key Laboratory of Quantum Information and Quantum Optics,
College of Physics and Information Engineering,
Fuzhou University, Fuzhou, Fujian 350108, China}

\author{Sheng Yang}
\altaffiliation{The first two authors contributed equally.}
\affiliation{Institute for Advanced Study in Physics and School of Physics, Zhejiang University, Hangzhou 310058, China}

\author{Shuo Liu}
\email{ls20@mails.tsinghua.edu.cn}
\affiliation{Institute for Advanced Study, Tsinghua University, Beijing 100084, China}

\author{Hai-Qing Lin}
\affiliation{Institute for Advanced Study in Physics and School of Physics, Zhejiang University, Hangzhou 310058, China}

\author{Shao-Kai Jian}
\email{sjian@tulane.edu}
\affiliation{Department of Physics and Engineering Physics, Tulane University, New Orleans, Louisiana, 70118, USA}

\begin{abstract}
Measurement-only quantum circuits offer a versatile platform for realizing intriguing quantum phases of matter. However, gapless symmetry-protected topological (gSPT) states remain insufficiently explored in these settings. In this Letter, we generalize the notion of gSPT to the critical steady state by investigating measurement-only circuits.  
Using large-scale Clifford circuit simulations, we investigate the steady-state phase diagram across several families of measurement-only circuits that exhibit topological nontrivial edge states at criticality. 
In the Ising cluster circuits, we uncover a symmetry-enriched non-unitary critical point, termed symmetry-enriched percolation, characterized by both topologically nontrivial edge states and string operator. 
Additionally, we demonstrate the realization of a steady-state gSPT phase in a $\mathbb Z_4$ circuit model. 
This phase features topological edge modes and persists within steady-state critical phases under symmetry-preserving perturbations. 
Furthermore, we provide a unified theoretical framework by mapping the system to the Majorana loop model, offering deeper insights into the underlying mechanisms. 
\end{abstract}

\maketitle

\emph{Introduction.}---A modern frontier in quantum many-body physics involves investigating how exotic quantum states can emerge in non-equilibrium settings. These investigations are of fundamental interest and also related to quantum simulation experiments~\cite{georgescu2014rmp,blatt2012quantum,altman2021prxquantum,noel2022measurement}. 
A prominent platform for investigating such phenomena is measurement-only quantum circuits incorporating non-commutative measurements~\cite{fisher2023random,xiang2013rmp}. The competing measurements introduce a novel form of frustration, enabling the realization of quantum steady states characterized by distinct orders~\cite{sang2021prr,BAO2021168618} and entanglement patterns~\cite{lavasani2021measurement,lavasani2021prl,lavasani2023prb,zhu2023structuredvolumelawentanglementinteracting,klocke2024entanglementdynamicsmonitoredkitaev,morral2023prb,klocke2022prb,kuno2023prb,sriram2023prb,orito2024prb,kuno2022emergencesymmetryprotectedtopological,zhang2024longrangeentanglementspontaneousnononsite,sukeno2024prb,lu2023prxquantum,lu2022prxquantum}, 
as well as the transitions between them~\cite{li20218prb,li2019prb,jian2020prb,vasseur2019prb,skinner2019prx,choi2020prl,jian2021prl,bao2020prb,turkeshi2021prb,lang2020prb,ippoliti2021prx,buchhold2021prx,tang2020prr,muller2022prl,minato2022prl,nahum2021prxquantum,tikhanovskaya2024prb,lu2024prb,turkeshi2022prb,liu2024prb,lu2021prxquantum,wang2024drivencriticaldynamicsmeasurementinduced,negari2024prb,liu2024prl,liu2023prb,klocke2023prx,nahum2020prr,nahum2013prb,sang2021prxquantum,zhang2022universal,zhang2021emergent,Jian2021prb, PhysRevB.107.094309, qian2024protectmeasurementinducedphasetransition}.

On a different front, recent advancements~\cite{cheng2011prb,fidkowski2011prb,kestner2011prb,keselman2015prb,ruhman2017prb,parker2018prb,JIANG2018753,keselman2018prb,scaffidi2017prx,thorngren2021prb,verresen2021prx,verresen2020topologyedgestatessurvive,DuquePRB2021,yu2022prl,yu2024prl,parker2019prl,yu2024prb,yang2024giftslongrangeinteractionemergent,zhong2024pra,umberto2021sci_post,friedman2022prb,li2023intrinsicallypurelygaplesssptnoninvertibleduality,huang2023topologicalholographyquantumcriticality,wen2023prb,wen2023classification11dgaplesssymmetry,wen2024stringcondensationtopologicalholography,li2024sci_post,huang2024fermionicquantumcriticalitylens,su2024prb,zhang2024pra,ando2024gaugetheorymixedstate,zhou2024floquetenrichednontrivialtopologyquantum,li2024noninvertiblesymmetryenrichedquantumcritical} have revealed that gapless quantum critical systems can support robust topological edge modes alongside critical bulk fluctuations~\cite{wen2017rmp,gu2009prb,chen2011prb,chen2012symmetry}. This phenomenon, known as symmetry-enriched quantum criticality or gapless symmetry-protected topological (gSPT) phases ~\cite{scaffidi2017prx,verresen2021prx,thorngren2021prb}, is characterized by topological edge modes~\cite{verresen2021prx}, nontrivial conformal boundary conditions~\cite{yu2022prl,parker2018prb}, and a universal bulk-boundary correspondence encoded in the entanglement spectrum~\cite{yu2024prl,zhang2024pra}. %Additionally, the discovery of nontrivial topology in critical systems suggests that topology plays a crucial role in classifying phase transitions even within the same universality class.% opening new avenues for understanding quantum phase transitions from a topological perspective. % and fundamentally changing the paradigm of phase transition research.

\begin{figure}[h]
    \centering
\includegraphics[width=0.8\linewidth]{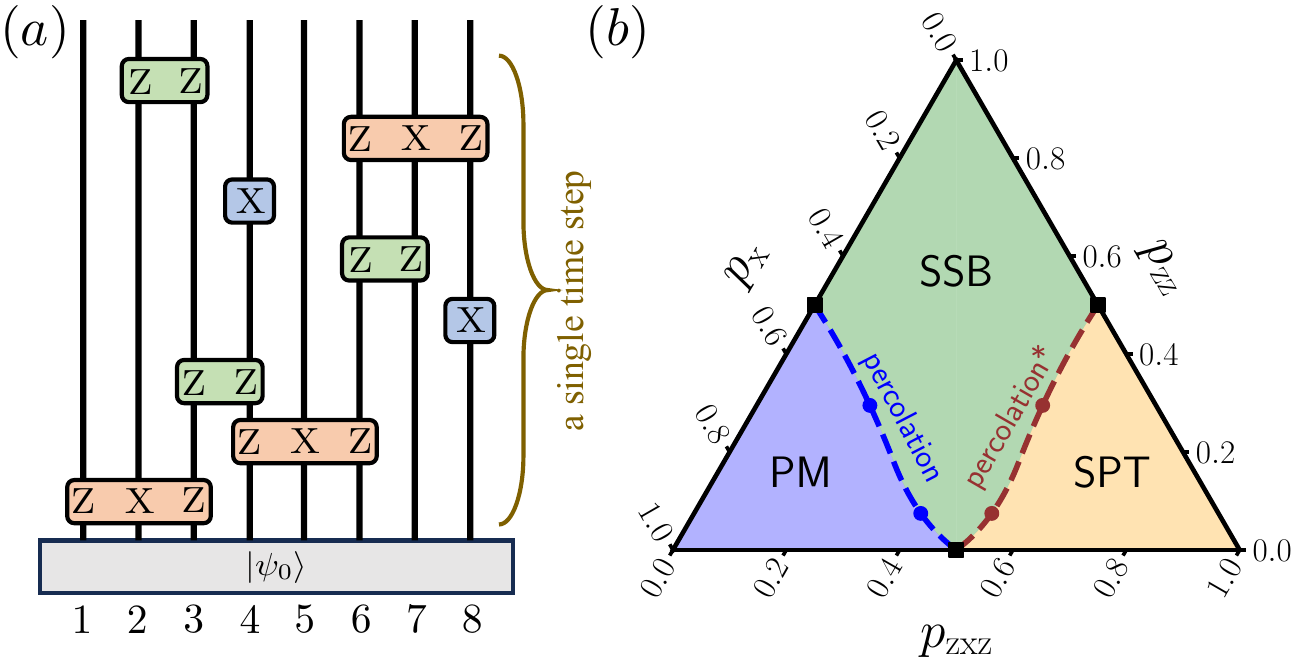}
    \caption{(a) Circuit diagram of the Ising cluster circuit model with $L=8$ qubits and $8$ measurements (i.e., a single time step): blue, green, and orange rectangles represent the projective measurements $X_{i}$, $Z_{i}Z_{i+1}$, and $Z_{i-1}X_{i}Z_{i+1}$, respectively.
    (b) The steady-state phase diagram of the cluster circuit as a function of the probabilities $p_{\text{x}}$, $p_{\text{zz}}$, and $p_{\text{zxz}}$. 
    The blue, green, and orange filled regions exhibit PM, SSB, and SPT orders, respectively. 
    %The black squares are exact critical points when only two types of the measurements are present. 
    The red circles are critical points obtained by the data collapse of the generalized topological entanglement entropy [see Fig.~\ref{fig:fig2}(a) and Sec.~II of the SM] and the red dashed line is a guide to the eye. Percolation$^*$ means the symmetry-enriched percolation universality.  The blue circles and the corresponding dashed line are obtained by $p_{\rm x} \leftrightarrow p_{\rm zxz}$.}
    \label{fig:setup}
\end{figure}

However, realizing gSPT phases in solid-state materials remains a significant challenge, highlighting the potential of quantum simulators as a promising platform for achieving these exotic quantum critical states.
This naturally raises the question: can the notion of gSPT be generalized to non-equilibrium settings, such as measurement-only circuits? 
If so, how can the underlying mechanisms behind these phenomena be analytically understood? 
To make progress in answering these questions, we investigate various families of measurement-only quantum circuits designed to generalize the notion of gSPT states to non-equilibrium settings. 
We firstly study the transition between different dynamical phases
and reveal a new type of universality, termed \emph{symmetry-enriched percolation}, featuring nontrivial boundary states. 
By generalizing and investigating the string operator with nontrivial symmetry flux, we find that the symmetry-enriched percolation cannot be connected to the conventional percolation without going through another fixed point, which in our case can be as a double-copy of percolation. 
Beyond critical points, we further generalize the gSPT phases to non-equilibrium settings. 
Specifically, we study the steady state phase diagram of a $\mathbb Z_4$ circuit, in which a gSPT phase that features nontrivial edge states as well as critical fluctuations presents over a significant portion of the phase diagram. 
The transitions to non-topological phases, including percolation and Berezinskii–Kosterlitz–Thouless (BKT) transition, are also unbiasedly identified.
Theoretically, the steady-state phase diagram can be understood by mapping the circuit model onto a Majorana loop model, providing a unified framework for investigating steady-state gSPT phases in 1+1D measurement-only circuits.

{\it Setup.}---We study two families of 1+1D measurement-only circuits, schematically illustrated in Fig.~\ref{fig:setup}(a). The circuit architecture consists of measurements randomly applied with certain probabilities. 
These measurements are uniformly selected from possible locations along a one-dimensional qubit chain of length $L$ under open boundary conditions (OBCs). 
The measurement protocol in this circuit is arranged as follows: 
A single-time step is defined as the application of $L$ random measurement operations during the time evolution. 
Each measurement operator is randomly selected from a predefined set of operators according to a specified probability. 
Starting from an initial state $\ket{\psi_{0}}$, %(e.g., a product state $\ket{+}^{\otimes L}$, where $X \ket{+} = \ket{+}$), 
we evolve it over a large number of time steps (set to $5L$ unless otherwise specified) to reach a steady state. Subsequently, we compute the target physical quantities (defined in detail in Sec.~I of the Supplementary Materials (SM)) and then average over different circuit realizations. 
% This protocol enables us to construct a steady-state phase diagram in the measurement-only setting.

\begin{figure}[htp]
    \centering
    \includegraphics[width=0.85\linewidth]{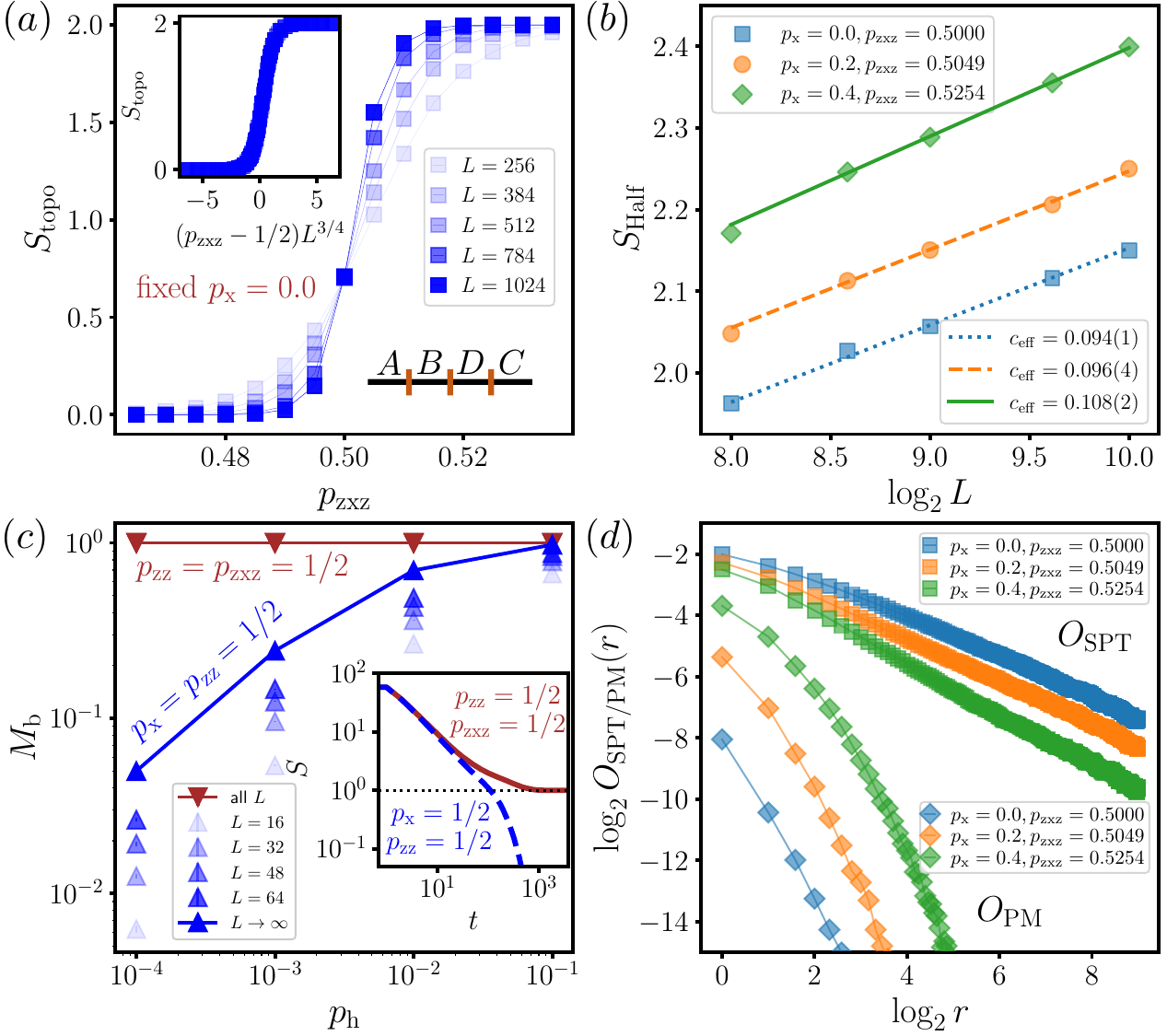}
    \caption{(a) The generalized topological entanglement entropy $S_{\rm topo}$ versus $p_{\rm zxz}$ for different system sizes with fixed $p_{\text{x}} = 0$. $A|B|D|C$ represents the equal partition used in the definition of $S_{\rm topo}$. The inset shows the data collapse of $S_{\text{topo}}$ with  $\nu = 4/3$ and $p_{\text{zxz},c} = 1/2$\,.
    (b) The half-chain entanglement entropy $S_{\text{Half}}$ grows logarithmically as the system size increases at the critical points for the case of $p_{\text{x}} = 0.0$, $0.2$, and $0.4$, respectively. The effective central charge $c_\text{eff}$ are obtained from least-squares fittings.
    (c) The edge magnetization, $M_{\rm b} \equiv (\overline{|\langle Z_{1} \rangle|} + \overline{|\langle Z_{L} \rangle|}) / 2$, in the presence of a small probability $p_{\rm h}$ of the boundary measurement $Z_{1/L}$ for the topologically trivial and nontrivial critical points, respectively. The inset shows the corresponding purification dynamics with an initial maximally mixed state for the same two critical points.
    (d) The string operators $O_{\rm SPT}(r)$ (square markers) and $O_{\rm PM}(r)$ (diamond markers) as a function of the lattice distance $r$ at the critical points for the case of $p_{\text{x}} = 0.0$, $0.2$, and $0.4$\,.}
    \label{fig:fig2}
\end{figure}

\emph{Symmetry-enriched percolation.}---We consider a $\mathbb{Z}_2$ symmetric Ising cluster model defined by a set of measurement operators, $\{ Z_{i-1}X_{i}Z_{i+1},  Z_i Z_{i+1}, X_i \}$, with corresponding probabilities $p_{\text{ZXZ}}$, $ p_{\text{ZZ}} $ and $ p_{\text{X}} = 1 - p_{\text{ZZ}} - p_{\text{ZXZ}}$, respectively. 
The equilibrium counterpart of this model features a ground state phase diagram with three phases: ferromagnetic spontaneous-symmetry-breaking (SSB), trivial paramagnetic (PM), and symmetry-protected topological (SPT) phases. Importantly, while the SSB-PM and SSB-SPT transitions are both described by the Ising conformal field theory (CFT), the time-reversal symmetry acts differently on the disorder operator~\cite{verresen2021prx,DuquePRB2021}, leading to distinct symmetry-enriched quantum critical points (QCPs). 
%This means the FM-SPT transition point hosts topologically protected edge modes coexisting with critical bulk fluctuations and cannot be smoothly deformed into the FM-PM transition even if they belong to the same universality class. 

In the measurement-only protocol described above, the ground-state phase diagram is replaced by a \emph{steady-state} phase diagram, as shown in Fig.~\ref{fig:setup}(b).
It exhibits the non-equilibrium analogs of SSB (spin-glass order), trivial PM, and SPT order~\cite{lavasani2021measurement,lavasani2021prl,sang2021prr,morral2023prb}. % (the complete steady-state phase diagram is presented in Sec.~II of the SM). %In the main text, we focus mainly on the critical and topological properties of two distinct dynamical critical points: $p_{\text{ZZ}} = p_{\text{ZXZ}} = 0.5$ and $p_{\text{ZZ}} = p_{\text{X}} = 0.5$ (see Sec.~II of the SM for the critical lines). 
The SPT (PM) phase emerges when $p_\text{ZXZ}$ ($p_\text{X}$) dominates. 
These phases are separated by an SSB phase when $p_\text{ZZ}$ is nonzero.  
We show that a nontrivial symmetry-enriched QCP emerges at the SSB-SPT transition at $p_{\text{ZZ}} = p_{\text{ZXZ}} = 1/2$, in stark contrast to the SSB-PM transition at $p_{\text{ZZ}} = p_{\text{X}} = 1/2$. 
To this end, we consider topological entanglement entropy by partitioning the whole system into four subregions with equal size, $A|B|D|C$.
$S_{\rm topo}$ is defined by $S_{\text{topo}} \equiv S_{AB}+S_{BC}-S_{B}-S_{ABC}$, where $S_A$ denotes the entanglement entropy of a subsystem $A$. 
In Fig.~\ref{fig:fig2}(a), by examining 
% the topological entanglement entropy 
$S_{\text{topo}}$, the critical point is located precisely at $p_{\text{ZZ}} = 1/2$, with a critical exponent $\nu = 4/3$. %which is attributed to self-duality~\cite{lavasani2021measurement,morral2023prb}. 
Furthermore, the half-chain entanglement entropy $S_{\text{Half}}$ exhibits an effective central charge $c_{\text{eff}} = 0.094(1) \approx \frac{\sqrt{3}\text{ln} 2}{4\pi}$ at the critical point, as shown in Fig.~\ref{fig:fig2}(b).
These behaviors unambiguously demonstrate that the SSB-SPT transition belongs to the bond percolation universality class~\cite{cardy1992critical,cardy2001conformalinvariancepercolation}. 
While the SSB-PM transition also belongs to the percolation class~\cite{lang2020prb,sang2021prr,sang2021prxquantum}, they are topologically distinct. 
To reveal the topological edge modes at steady-state criticality, we calculate the edge magnetization $M_{b}$ as a function of the boundary field probability $p_h$ across different system sizes under OBCs in Fig.~\ref{fig:fig2}(c). 
At the SSB-PM transition, the edge magnetization decreases with decreasing $p_h$ (blue triangles), indicating the absence of edge modes.  
In contrast, at the SSB-SPT transition, the edge magnetization remains finite (red solid line) as $p_h$ approaches zero, providing direct evidence of topological edge modes at criticality.
Moreover, in the inset, we show the full entanglement entropy under a purification dynamics, with the initial state being the maximally mixed state.  
The SSB-SPT (SSB-PM) transition shows a nonvanishing (vanishing) residue entropy $S(t)=1$ ($S(t) = 0$) at late time $t \rightarrow \infty$, indicating the topological edge states.

The distinct topological properties suggest that these two QCPs cannot be adiabatically connected without going through another critical point. 
The phase diagram in Fig.~\ref{fig:setup}(b) shows a generic path connecting the two QCPs, where the SSB-SPT (SSB-PM) transition is denoted respectively by the red (blue) dashed curves. 
The effective central charge is shown to be the same along these two transitions, as shown in Fig.~\ref{fig:fig2}(b), indicating that they both belong to the same percolation class;
hence, it is crucial to show the distinction between the two QCPs along the path, as well as the emergence of another critical point. %,which occurs at the SPT-PM transition with $p_\text{X} = p_\text{ZXZ} = 1/2$~\cite{}. 
To this end, we use the following string operators~\cite{verresen2021prx,morral2023prb}: 
\begin{eqnarray}
\label{eq:orders}
    O_{\rm PM}(i,j) & =& \overline{|\langle X_{i} X_{i+1} \cdots X_{j-1} X_{j} \rangle|} \,, \\
    O_{\rm SPT}(i,j) & =&  \overline{|\langle Z_{i-1} Y_{i} X_{i+1} \cdots X_{j-1} Y_{j} Z_{j+1} \rangle|} \,,
\end{eqnarray}
where $\langle \cdot \rangle $ denotes the expectation value taken in the steady state and $\bar{\cdot}$ denotes the average over steady states for different circuit realizations.
Absolute value is also required to achieve a nonvanishing result~\cite{morral2023prb}. 
Importantly, the SPT string operator $O_{\rm SPT}(i,j)$ carries a nontrivial symmetry flux that is odd under time-reversal symmetry at the endpoints. 

Figure~\ref{fig:fig2}(d) 
%We further analyze various types of string order parameters as a function of lattice distance at criticality, demonstrating that string order parameter $O_{\text{SPT}}$ can be viewed as a symmetry-flux operator in the sense of symmetry-enriched quantum criticality~\cite{verresen2021prx} (see Sec.~II of the SM for details), resulting in topologically edge modes even in the steady state. 
shows the SPT string operator dominates at the symmetry-enriched QCP along the SSB-SPT transition, whereas the PM string operator dominates at the SSB-PM transition (see SM Sec.~II~D). 
Since the two string operators carry distinct symmetry fluxes, they cannot be smoothly changed without another critical point. 
Indeed, the SPT-PM transition point at $p_\text{X} = p_\text{ZXZ} = 1/2$ is a different QCP with an effective central charge $c_{\text{eff}} =2 \times \frac{\sqrt{3}\text{ln} 2}{4\pi}$ that connects the symmetry-enriched SSB-SPT QCP and the SSB-PM QCP. 
Notice that the SPT-PM transition can be understood by two copies of percolation~\cite{lavasani2021measurement}. 
Therefore, our results reveal the emergence of \emph{symmetry-enriched percolation} in the measurement-only circuit, making the first example of symmetry-enriched non-unitary CFT. 
Note that, in the Appendix, we use the Majorana loop model to provide a theoretical understanding of the symmetry-enriched QCP in this model.

\emph{gSPT phase in the steady state.}---To extend the notion of topology to dynamical \emph{critical phases}, we investigate the steady state of a $\mathbb{Z}_{4}$-symmetric measurement-only circuit defined by a measurement operator set consisting of five different types of operators: 
The first three types are inspired by the equilibrium intrinsic gSPT model~\cite{li2023intrinsicallypurelygaplesssptnoninvertibleduality,li2024sci_post} (briefly reviewed in the SM Sec.~III~A), $\{\tau_{2i-1}^{z} \sigma_{2i}^{x} \tau_{2i+1}^{z}, \tau_{2i-1}^{y}\sigma_{2i}^{x}\tau_{2i+1}^{y}, \sigma_{2i}^{z} \tau_{2i+1}^{x}\sigma_{2i+2}^{z} \}$ with an equal probability $p_t$, and the last two types are competing measurements, $\{ \tau_{2i-1}^{x}\tau_{2i+1}^{x}, \sigma^x_{2i}\}$
with corresponding probabilities, $\{  p_\delta, p_h \}$. 
Note that $3p_t + p_\delta + p_h= 1$. 
Here, each pair of $(\tau_{2i-1},\sigma_{2i})$ represents the $i$th unit cell, and the two species of spins per unit cell are represented by Pauli operators $\sigma^{\alpha}$ and $\tau^{\alpha}$. 
The operators possess a $\mathbb{Z}_{4}$ symmetry defined by $U=\prod_{i}\sigma_{2i}^{x} e^{i\frac{\pi}{4}(1-\tau_{2i-1}^{x})}$. 
The equilibrium counterpart exhibits an intrinsic gapless SPT state when the first three operators dominate. 
This phase is protected by an emergent anomaly of the $\mathbb Z_4$ symmetry. %, and stable against the perturbation of the last two operators.

\begin{figure}
    \centering
    \includegraphics[width=0.85\linewidth]{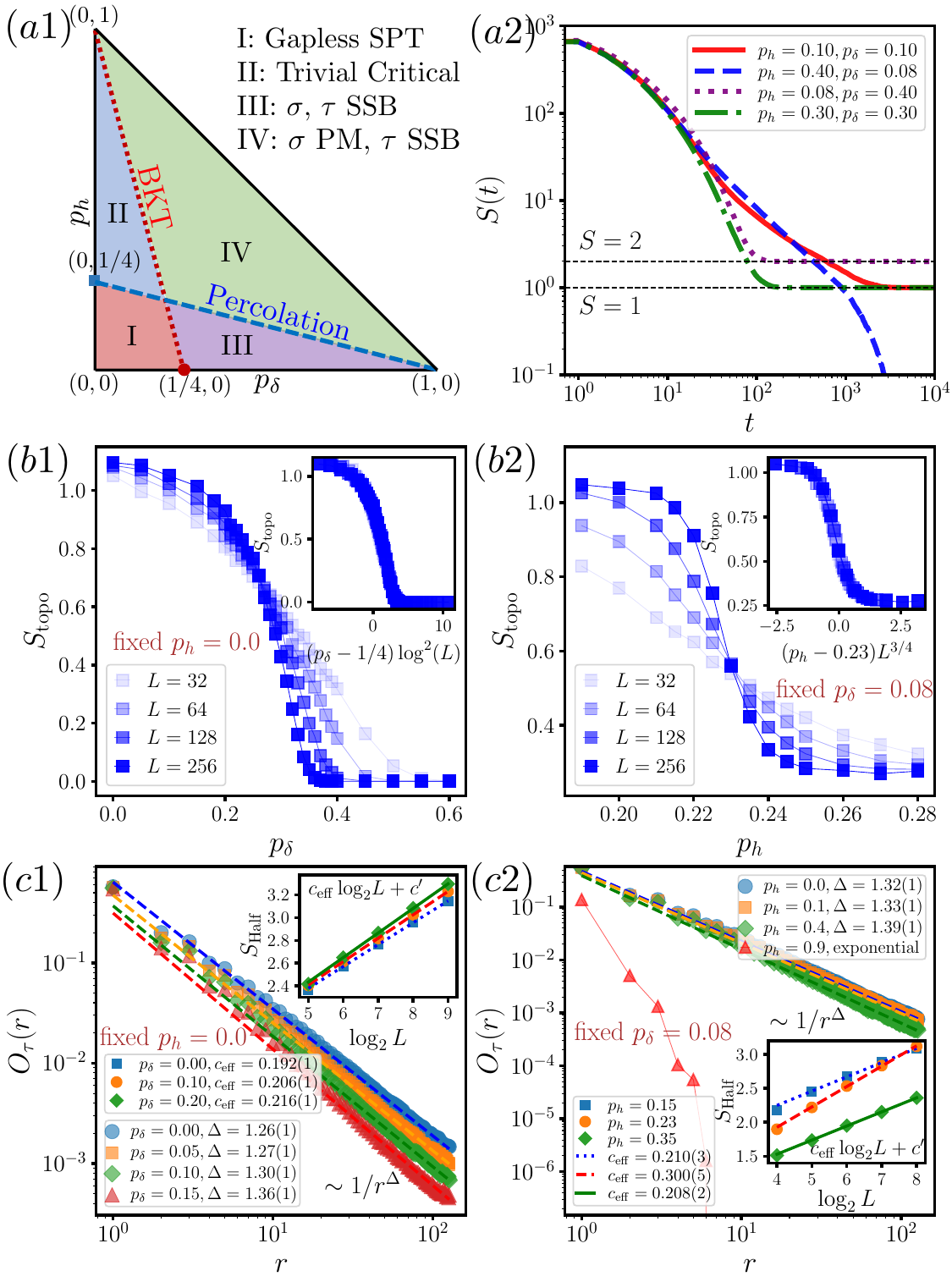}
    \caption{(a1) Steady state phase diagram of the $\mathbb Z_4$ symmetric circuits. The BKT (percolation) transition line is $4p_{\delta} + p_{h} = 1$ ($4p_{h} + p_{\delta} = 1$) (see Sec.~III~B of the SM for additional numerical results).
    (a2) The evolution of the full system entropy, $S$, starting from a maximally mixed state for four representative points in different phases.
     $S_\text{topo}$ as a function of $p_{\delta}$ at $p_{h} = 0$ (b1) and as a function of $p_{h}$ at $p_{\delta} = 0.08$ (b2). The inset gives the data collapse of $S_\text{topo}$. %indicating a BKT transition at the critical point $p_{\delta} = 0.25$\,.
     %A similar analysis indicates a percolation transition at the critical point $p_{h} = 0.23$ with $\nu = 4/3$ when $p_{\delta} = 0.08$ is fixed.
    (c1) The string operator $O_{\tau}(r)$ exhibits a power-law decay when $p_{\delta} \in [0, 0.25)$ for $p_{h} = 0$. Data within the interval $r \in (L/8, L/2)$ are used in the fittings and one end of the string operator was fixed at the boundary. The inset shows $S_{\rm Half}$ as a function of $L$.
    (c2) The string operator $O_{\tau}(r)$ exhibits a power-law dependence on the lattice distance in the Phase~I and Phase~II, while an exponential decay in the Phase~IV. The inset shows the logarithmic behavior of $S_{\rm Half}$ as a function of $L$ for three representative points. 
    The effective central charge $c_\text{eff}$ is obtained from the least-squares fitting. The simulated system size is $L = 512$ in (a2) and $L = 256$ in (c1) and (c2).}
    \label{fig:fig3}
\end{figure}

Since the emergent anomaly is absent in the context of measurement-only circuits, where energy conservation does not apply, whether gSPT phases still exist in such settings is an outstanding question. 
We reveal an intriguing steady state phase diagram in Fig.~\ref{fig:fig3} (a1), in which Phase I represents a gSPT phase with nontrivial edge states characterized by a nonvanishing residue entropy in the purification dynamics, nontrivial topological entanglement entropy, as well as critical correlation functions detailed below.  
Specially, Fig.~\ref{fig:fig3}(a2) shows the entanglement entropy under a purification dynamics:   
The red curve with a nontrivial residue $S= 1$ entropy indicates two degenerate states. 
Moreover, with nontrivial topological entanglement entropy at small but finite $p_\delta$ and $p_h$,  Fig.~\ref{fig:fig3}(b1, b2) shows that the degeneracy in Phase I is originated from topological edge states. 
Finally, we can show that the steady state is also gapless by examining the string operator, 
$O_{\rm \tau}(|i-j|) = \overline{| \langle \tau_{2i-1}^{z} (\prod_{k=i}^{j-1} \sigma_{2k}^{x}) \tau_{2j-1}^{z} \rangle |}$,
and the half-chain entanglement entropy. 
Figure~\ref{fig:fig3}(c1) shows a power law behavior of the string operator $O_\tau$ with an exponent $\Delta \approx 4/3$, and an effective central charge $c_{\text{eff}} \approx 2 \times \frac{\sqrt{3}\text{ln} 2}{4\pi}$ in the inset. 
Later, we will see that the critical state is equivalent to two copies of percolation, which fully explains the observed $c_\text{eff}$. 

Now let's discuss the transition to Phase II and III due to the competing measurement operators. 
The Phase III is an SSB state for both $\sigma$ and $\tau$ degrees of freedom, induced by the perturbation $\tau^x_{2i-1} \tau^x_{2i+1}$. 
The residue entanglement entropy $S(t) =2$ at late time in the purification dynamics is originated from the SSB for both $\sigma$ and $\tau$ degrees of freedom, as shown by the dotted purple curve in Fig.~\ref{fig:fig3}(a2). 
The two-point functions of $\tau^x $ and $\sigma^z$ both develop long-range correlation due to the SSB (detailed in the SM Sec.~III~C). 
The transition between the gSPT phase and the SSB phase belongs to the BKT universality class, as unveiled by a perfect data collapse of $S_{\rm topo}$ in Fig.~\ref{fig:fig3}(b1), which confirms a logarithmic-squared scaling form near the critical point~\cite{klocke2023prx}. 
Note that the topological entropy also decreases to zero due to the absence of topological edge state.
On the other hand, the competing measurement $\sigma^x_{2i}$ leads to a transition to Phase II, which is a trivial critical phase without topological edge state. 
The string operator exhibits the same power law and the half-chain entanglement entropy also gives the same effective central charge, as shown in Fig.~\ref{fig:fig3}(c2) indicating the critical phase is of the same nature. 
However, the residue entropy, shown by the blue dashed curve in Fig.~\ref{fig:fig3}(a2), and the topological entropy in Fig.~\ref{fig:fig3}(b2) both vanish, demonstrating that Phase II is a non-topological gapless state.
In Fig.~\ref{fig:fig3}(b2), we further reveal that the transition belongs to the percolation transition with the exponent $\nu = 4/3$. 
Lastly, Phase IV is an SSB phase for the $\tau$ degrees of freedom \footnote{It can be understood by a symmetry breaking transition for $\tau$ spins from the Phase II due to the $\tau^x_{2i-1} \tau^x_{2i+1}$ perturbation,
or by a symmetry restoration for $\sigma$ spins from the Phase III due to the increase in the $\sigma^x$ measurements.}.

%Although the topological critical phase in the steady state usually lacks the emergent anomaly and is therefore no longer an intrinsically gSPT phase (we can construct the intrinsically gSPT in a steady state by postselection). 
%These topological critical steady phases remain robust against symmetry-preserving two-site perturbations but transition into $\mathbb{Z}_{2}$ symmetry-breaking area-law phases upon adding a symmetry-preserving single-site perturbation. These phases occupy a larger parameter space in the steady-state phase diagram (see Fig.~\ref{fig:setup}(b) and Sec.~III of the SM for additional numerical evidence), resulting in an unexpected qualitatively distinct phase structure compared to the ground-state phase diagram~\cite{kuno2023prb}. In the next section, we map the circuit model to a Majorana loop model to provide further insight into the above numerical observations.
\begin{figure}
    \centering
    \includegraphics[width=0.8\linewidth]{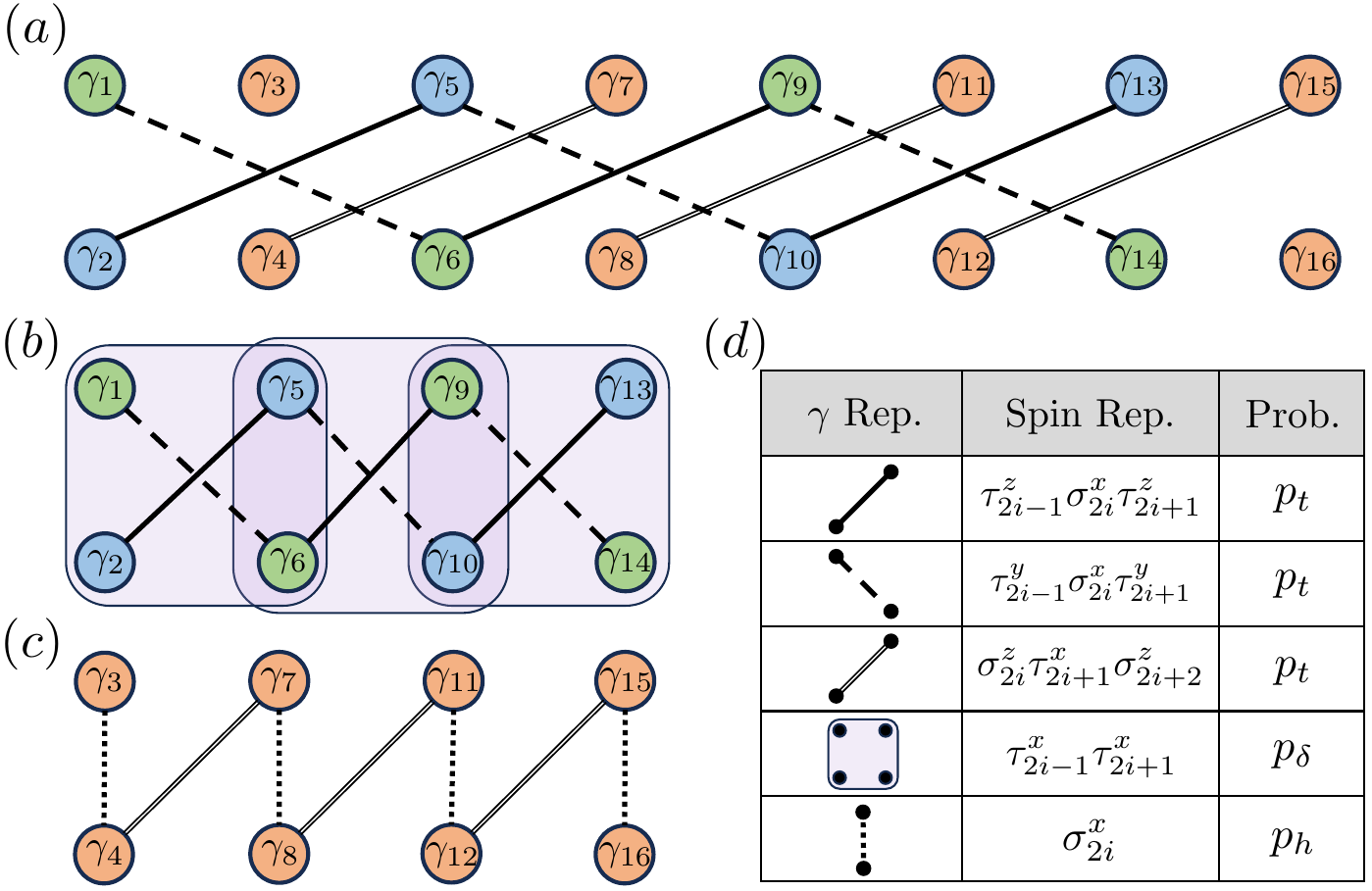}
    \caption{Majorana representation of the $\mathbb{Z}_{4}$ symmetric circuit.
    (a) When $p_{\delta} = p_{h} = 0$, there are only three-site measurements \{$\tau_{2i-1}^{z}\sigma_{2i}^{x}\tau_{2i+1}^{z}$, $\tau_{2i-1}^{y}\sigma_{2i}^{x}\tau_{2i+1}^{y}$, $\sigma_{2i}^{z}\tau_{2i+1}^{x}\sigma_{2i+2}^{z}$\}; this model can be seen as two decoupled Majorana chains exhibited by (b) and (c).
    (b) The two-site perturbation $\tau_{2i-1}^{x}\tau_{2i+1}^{x}$ is added on the $\tau$ degrees of freedom, which is a four-Majorana operator.
    (c) The single-site perturbation $\sigma_{2i}^{x}$ is added on the $\sigma$ degrees of freedom.
    (d) gives the table of the associated projective measurements and their corresponding probabilities and Majorana representations.}
    \label{fig:fig4}
\end{figure}

\emph{Majorana loop models for measurement-only circuits.}---We present a framework for the measurement-only circuit model based on the Majorana loop model to understand the numerical results. 
Here, we focus on the $\mathbb Z_4$ symmetric circuit. 
See the Appendix for more details.

First, consider the case without perturbation, i.e., $p_{h} = p_{\delta} = 0$. After performing the Jordan-Wigner transformation, the three-site measurements, $\tau_{2i-1}^{z}\sigma_{2i}^{x}\tau_{2i+1}^{z}$, $\tau_{2i-1}^{y}\sigma_{2i}^{x}\tau_{2i+1}^{y}$, and $\sigma_{2i}^{z}\tau_{2i+1}^{x}\sigma_{2i+2}^{z}$, are mapped to Majorana parity measurements, as shown in Fig.~\ref{fig:fig4}(a), on $(4i-2, 4i+1)$, $(4i-3, 4i+2)$ and $(4i,4i+3)$, respectively. 
In the Majorana representation, the whole system becomes two decoupled parts: $\tau$-chain and $\sigma$-chain [see also Figs.~\ref{fig:fig4}(b) and~\ref{fig:fig4}(c) respectively].
It becomes evident that the edge modes are represented by the two dangling Majorana fermions in the $\sigma$-chain. 
In the absence of perturbation, the Majorana modes in the $\tau$-chain can be further divided into two independent sets as shown in Fig.~\ref{fig:fig4}(b), which corresponds to two percolation models~\cite{klocke2023prx}, whereas, the $\sigma$-chain is noncritical due to the dimerized pattern.
Hence, the Majorana representation clearly unveils the gSPT steady state without perturbation.

Naively, the decoupled $\sigma$-chain resembles the Majorana representation of an Ising model, in which the two edge states account for the two-fold degeneracy from SSB.
However, in terms of the spin operators in the $\mathbb Z_4$ circuit, the $\sigma$ spin and the $\tau$ spin are strongly coupled. The gapless fluctuation in the $\tau$-chain renders the correlation of $\sigma$ spins gapless and disorders the seemingly SSB phase. 
This can be seen by noticing that the correlation of local spin operators $\sigma^z_{2i} \sigma^z_{2j}$ overlaps with the gapless Majorana in the $\tau$-chain, preempting the SSB correlation.
Instead, only the string operator $\sigma_{2i}^{z} (\prod_{k=i}^{j-1} \tau_{2k+1}^{x}) \sigma_{2j}^{z} $ purely overlaps with the dimerized Majorana, leading to a long-range correlation. 
In this sense, the topological edge state is protected by the gaplessness, and indeed, we will see in the following that once the gapless Majorana in the $\tau$-chain is gapped out, the local spin operators $\sigma^z_{2i} \sigma^z_{2j}$ acquire a long-range order, rendering the state non-topological.

The two-site measurement perturbation $\tau_{2i-1}^{x}\tau_{2i+1}^{x}$ corresponds to a non-Gaussian Majorana measurement as shown in Fig.~\ref{fig:fig4}(b) and induces couplings between the two percolation models (two $\tau$-chains in the unperturbed circuit). 
When $p_{\delta}$ is small, these couplings are irrelevant~\cite{Fendley_2008, klocke2023prx}, the gSPT phase remains robust. 
When $p_{\delta}$ is large enough, the couplings between the two percolation chains become relevant and lead to a BKT transition to a noncritical state corresponding to the $\tau$ SSB phase in the spin picture.
Moreover, the topological edge state from $\sigma$ spins loses the protection and develops a long-range order.
This is fully consistent with the $\sigma$ and $\tau$ SSB in the Phase III. 

Since the topological edge state is protected by the gapless fluctuations, relevant deformations that gap out the state will also render the state non-topological.
For instance, unlike the equilibrium counterpart, the single-site operator $\tau^x_{2j+1}$ is a relevant perturbation at the fixed point with two copies of percolation~\cite{klocke2023prx}. 
We also investigate the effect of such a relevant perturbation, and find that indeed it leads to a transition from the gSPT phase to a $\sigma$ SSB state (See SM Sec.~III~D for details). 
Of course, this perturbation can be forbidden by a $\mathbb Z_2$ symmetry, $\tau^x \rightarrow - \tau^x$. 

Finally, the single-site measurement perturbation, $\sigma_{2i}^{x}$, simply 
leads to the coupling on $(4i-1,4i)$, as indicated in Fig.~\ref{fig:fig4}(d), which is the other dimerization in Fig.~\ref{fig:fig4}(c). 
Since the $\sigma$ chain and the $\tau$ chain are decoupled, the percolation transition at the equal strength of both dimerizations is consistent with the transition to the Phase II.

\emph{Concluding remarks.}---To conclude, we uncover novel quantum phases and critical points in measurement-only circuits, including the first example of a symmetry-enriched non-unitary CFT and a steady-state gSPT phase with robust edge modes. By mapping these systems to the Majorana loop model, we also provide a unified framework for understanding these phenomena, highlighting measurement-only circuits as a powerful platform for exploring exotic quantum states and transitions.

\textit{Acknowledgement}: We thank Ruochen Ma for helpful discussions. X.-J. Yu was supported by the National Natural Science Foundation of China (Grant No.12405034). This work is also supported by MOST 2022YFA1402701. S. Y. was supported by China Postdoctoral Science Foundation (Certificate Number: 2024M752760).
The work of S.-K. J. is supported by a start-up grant and a COR Research Fellowship from Tulane University.

{\it Appendix: More details on the Majorana loop model and the cluster circuit model.---}In the following, we present the details of the mapping from the measurement-only circuits to the Majorana loop models~\cite{klocke2023prx} and provide the theoretical understanding of the Ising cluster circuit model in Fig.~\ref{fig:setup}.

To obtain the Majorana loop model, we first perform the Jordan-Wigner transformation to map the spin operators to the Majorana fermions,
\begin{equation}
    \gamma_{2j-1} = \bigg( \prod_{k=1}^{j-1} X_{k} \bigg) Z_{j} \,, \qquad{} \gamma_{2j} = \bigg( \prod_{k=1}^{j-1} X_{k} \bigg) Y_{j} \,,
\end{equation}
where $\gamma_{2j-1}$ and $\gamma_{2j}$ are two Majorana fermions at site $j$ and satisfy the anti-commutation relation $\{\gamma_{m}, \gamma_{n}\}_{+} = 2 \delta_{m,n}$. Subsequently, the projective measurements used in the measurement-only circuits can be mapped to the Majorana measurements as summarized in Fig.~\ref{fig:fig4} for the $\mathbb{Z}_{4}$ circuit model. For the Ising cluster circuit model, the three types of projective measurements correspond to the following Majorana measurements,
% As summarized in Fig.~\ref{fig:ci_majorana}, the three types of projective measurements in the cluster circuit model correspond to the following Majorana measurements
\begin{eqnarray}
    \label{eq:ci_measurement}
    &&X_{j} = i \gamma_{2j-1} \gamma_{2j} \,, 
    \qquad{} Z_{j}Z_{j+1} = i \gamma_{2j} \gamma_{2j+1} \,, \qquad{}  \\ \nonumber
    &&Z_{j-1}X_{j}Z_{j+1} = i \gamma_{2j-2} \gamma_{2j+1} \,,
\end{eqnarray}
where $i \gamma_{m} \gamma_{n}$ is the Majorana parity measurement on flavors $(m,n)$.  
As a result, the projector for positive and negative parity is $\frac{1\pm i \gamma_m \gamma_n}{2}$. As shown in Fig.~\ref{fig:ci_majorana}(a), we use a single arc connecting $\gamma_m$ and $\gamma_n$ to represent the Majorana measurement $i\gamma_m \gamma_n$ regardless of the outcomes. 
We note that the measurements in the Ising cluster circuit model are all Gaussian Majorana measurements, while the $\tau_{2i-1}^{x}\tau_{2i+1}^{x}$ measurement in the $\mathbb{Z}_{4}$ circuit model is a non-Gaussian Majorana measurement. 
After establishing the transformation between the projective measurements in the spin basis and the Majorana measurements, we proceed to introduce the mapping from the measurement-only circuit to the Majorana loop model~\cite{klocke2023prx} via the following steps:
\begin{itemize}
    \item[1).] Firstly, the initial state can be represented by a specific pairing configuration of the Majorana fermions. For the initial product state $\ket{\psi_{t=0}} = \bigotimes_{i=1}^{L} \ket{+}_{i}$ considered in the work, where $\ket{+}_{i}$ is the eigenstate of $X_{i}$ with $+1$ eigenvalue, it can be represented by Majorana fermions with a pairing configuration $\{ (\gamma_{1},\gamma_{2}), (\gamma_{3},\gamma_{4}), \cdots (\gamma_{2L-1},\gamma_{2L}) \}$.
    
    \item[2).] Subsequently, the Majorana parity measurements will rearrange the pairing configuration. For the Majorana fermions with pairing configuration $\{\cdots, (\gamma_{l}, \gamma_{m}), (\gamma_{n}, \gamma_{p})  ,\cdots \}$ at a discrete time step $t$, the parity measurement $i \gamma_{m} \gamma_{n}$ will break the original pairings  $(\gamma_{l}, \gamma_{m})$ and $(\gamma_{n}, \gamma_{p})$, and create new pairings $(\gamma_{l}, \gamma_{p})$ and $(\gamma_{m}, \gamma_{n})$. Consequently, the pairing configuration at discrete time step $t+1$ is $\{\cdots, (\gamma_{l}, \gamma_{p}), (\gamma_{m}, \gamma_{n})  ,\cdots \}$. Therefore, the dynamics of the measurement-only circuits can be described by the dynamics of the Majorana pairing configurations along the time evolution, i.e., Majorana fermion worldlines.
\end{itemize}
Based on the points listed above, the instantaneous quantum state at discrete time $t$ is determined by the corresponding Majorana pairing configuration (up to the sign of the parity), and thus the measurement-only circuits are mapped to the Majorana loop models. 
%Note that the associated Pauli measurements all correspond to Majorana parity measurements on two Majorana modes in the Ising cluster model. Therefore, there will be exactly $L$ Majorana pairs at any time. 

\begin{figure}
    \centering
    \includegraphics[width=\linewidth]{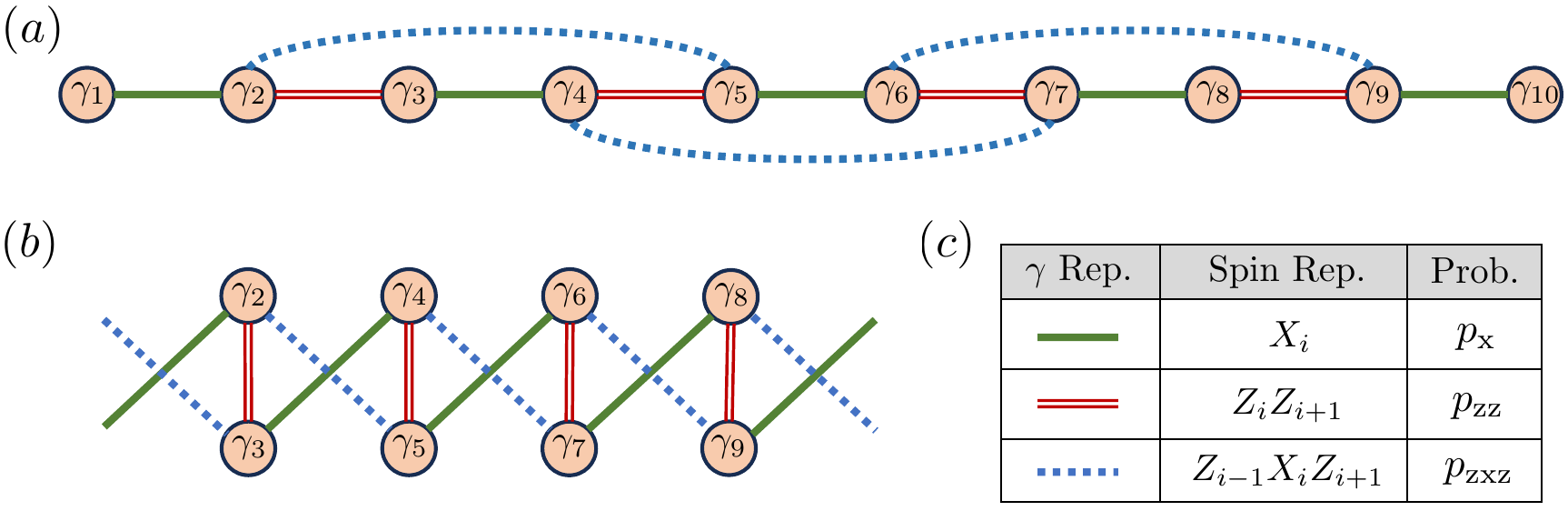}
    \caption{Majorana representation of the cluster circuit model composed of $X_{i}$, $Z_{i}Z_{i+1}$, and $Z_{i-1}X_{i}Z_{i+1}$ measurements. (a) Each spin on site $i$ is represented by two Majorana modes $\gamma_{2i-1}$ and $\gamma_{2i}$. The green solid line, the red double solid line, and the blue dotted line represent the one-site $X_{i}$, the two-site $Z_{i}Z_{i+1}$, and the three-site $Z_{i-1}X_{i}Z_{i+1}$ measurements in the Majorana representation respectively. (b) A rearrangement of (a) preserving the linking structure under periodic boundary conditions. It is easy to see that there is a symmetry $p_{\text{x}} \leftrightarrow p_{\text{zxz}}$. (c) gives the table of the associated projective measurements and their corresponding probabilities and Majorana representations.}
    \label{fig:ci_majorana}
\end{figure}

Our observables have a direct correspondence in the Majorana loop model. 
Consider the von Neumann entanglement entropy $S_{R}$, where $R$ is a contiguous system interval. In the Majorana loop model, it corresponds to the number of Majorana pairs, $N_{\rm pair}$, in which one Majorana fermion is in the region $R$ and the other is in the complement $\bar{R}$, and $S_{R}= N_{\rm pair}/2$~\cite{sang2021prxquantum, lang2020prb, fattal2004entanglement}. Furthermore, the properties of the ensemble-averaged $S_{R}$ are captured by the normalized arc length distribution $P(l)$ ($l = |m-n|$ is the pair length for the Majorana pair $(\gamma_{m}, \gamma_{n})$), which is in one-to-one correspondence with the stabilizer length distribution in the clipped gauge~\cite{li2019prb, PhysRevX.7.031016}. By tuning to the critical point, the ensemble of the Majorana loop model is described by the bond percolation conformal field theory (CFT) with critical exponent $\nu = 4/3$ and the length distribution $P(l)$ of the Majorana pairs obeys a universal form~\cite{Jacobsen2008PRL,nahum2013prb} $l^2 P(l) \approx \frac{\sqrt{3}}{2\pi}$. 
This universal constant is then related to the prefactor $c_\text{eff}$ in the logarithmic $S_{\rm Half}$ and $c_\text{eff} = \sqrt{3} \ln2 / (4\pi)$~\cite{klocke2023prx,sang2021prxquantum}. % under open boundary conditions.

The purification dynamics can also be understood from the perspective of the Majorana loop model. 
Unlike the case where the system starts from a product state with initial $L$ Majorana pairings, as discussed earlier, there are no pairings at $t=0$ when a maximally mixed state is chosen as the initial state in the purification dynamics. 
As time increases, Majorana fermion pairings are formed. The total entropy is $S(t)=\frac{1}{2} n_{s}(t)$, with $n_{s}(t)$ being the spanning number counting the unpaired Majorana modes. 
As detailed below, the distinct purification dynamics at SSB-PM and SSB-SPT transitions can be straightforwardly understood from the Majorana representations.

The Majorana loop model gives a straightforward understanding of the connection and distinction between the SSB-PM and SSB-SPT transitions. 
In the Majorana representation, the symmetry $p_{\text{x}} \leftrightarrow p_{\text{zxz}}$ used in Fig.~\ref{fig:setup} becomes transparent in Fig.~\ref{fig:ci_majorana}(b). Consequently, in the absence of $Z_{i}Z_{i+1}$ measurement, the SPT-PM transition occurs at $p_{\text{x}} = p_{\text{zxz}} = 1/2$. Moreover, as shown in Fig.~\ref{fig:ci_majorana}(b), it is obvious that the measurement-only circuit model corresponds to two decoupled Majorana chains and thus the critical point $p_{\text{x}} = p_{\text{zxz}} = 1/2$ is characterized by two copies of the critical percolation. 
% After a suitable rearrangement of the Majorana chain given in Fig.~\ref{fig:ci_majorana}(a) into a Majorana ladder as shown in Fig.~\ref{fig:ci_majorana}(b), we can easily see that the circuit model in the special case of $p_{\text{zz}} = 0$ can be seen as two decoupled Majorana chains, and the critical point $p_{\text{x}} = p_{\text{zxz}} = 1/2$ is characterized by two copies of the critical percolation. 
% Moreover, the symmetry $p_{\text{x}} \leftrightarrow p_{\text{zxz}}$ of the steady-state phase diagram [see Fig.~\ref{fig:order_diagram}] becomes more transparent in the Majorana loop representation as explained by Fig.~\ref{fig:ci_majorana}(b). 
However, it is also noted that the existence of the symmetry $p_{\text{x}} \leftrightarrow p_{\text{zxz}}$ does not mean that the two percolation lines related by this symmetry are completely equivalent. 
As shown in the main text and Section~II in the Supplemental Materials, the SSB-SPT transition is different from the SSB-PM transition under open boundary conditions as revealed by the boundary magnetization, non-trivial residue entropy, and the dominated $O_{\text{SPT}}(r)$. 
The difference comes from the fact that the physical degree of freedom on site $i$ is composed of the Majorana fermions, $\gamma_{2i-1}$ and $\gamma_{2i}$, rather than $\gamma_{2i-2}$ and $\gamma_{2i+1}$. 
Consequently, in the Majorana representation, there are two Majorana modes, $\gamma_{1}$ and $\gamma_{2L}$, unmeasured in the whole dynamics at the parameter point $p_{\text{zz}} = p_{\text{zxz}} = 1/2$ and $p_{\text{x}} = 0$ and induce the residue entropy $S=1$ in the purification dynamics. In the setup with initial product states, these two Majorana modes are paired in the final state and a robust long-range entanglement is established in the system. 

Further, the string operators in the Majorana representation ($m < n$ is assumed) read
\begin{eqnarray}
     \hat O_\text{PM} &=& \prod_{k=m}^{n} X_{k} \sim (i)^{n-m+1} \prod_{k=2m-1}^{2n} \gamma_{k} \,, \\
     %Z_{m} Z_{n} \sim (i)^{n-m} \prod_{k=2m}^{2n-1} \gamma_{k} \,, \\
     \hat O_\text{SPT} &=& Z_{m-1} Y_{m} \bigg( \prod_{k=m+1}^{n-1} X_{k} \bigg) Y_{n} Z_{n+1} \nonumber \\
     && \sim (i)^{n-m+3} \gamma_{2m-2} \bigg( \prod_{k=2m}^{2n-1} \gamma_{k} \bigg) \gamma_{2n+1} \,.
\end{eqnarray}
It is easier to consider SSB-PM transition with $p_\text{ZXZ}=0$, while the case at the SSB-SPT transition can be understood similarly by the symmetry. 
Since the PM string operator $\hat O_\text{PM}$ is a consecutive product of Majorana operators in an interval from $m$ to $n$, it is nonvanishing only when Majorana fermions inside the interval are paired within themselves. 
Equivalently, a configuration with a pairing between a Majorana inside the interval and another Majorana outside is not allowed. 
With the mapping to the one-state Potts model~\cite{temperley2004relations,cardy2001conformalinvariancepercolation}, this string operator corresponds to two boundary condition changing (bcc) operators located at the end points of the interval. 
This bcc operator has a scaling dimension $\Delta_\text{bcc} = 1/3$, leading to a power law behavior with exponent $2\Delta_\text{bcc} =2/3$ of the string operator $\hat O_\text{PM}$ at the SSB-PM transition.  
On the contrary, the SPT string operator $\hat O_\text{SPT}$ is not a consecutive product, and will decay exponentially at the SSB-PM transition.
The symmetry-enriched percolation at the SSB-SPT transition can be understood via the dual transformation of $p_\text{x} \leftrightarrow p_\text{zxz}$, so that the SPT string operator $\hat O_\text{SPT}$ features a power law behavior with exponent $2\Delta_\text{bcc} =2/3$ whereas the PM string operator $\hat O_\text{SPT}$ decays exponentially, which fully explains the numerical results in Fig.~\ref{fig:fig2}(d). 
%More specifically, the Majorana product shown in the Pauli operator $Z_{m}Z_{n}$ can be decomposed into the product of the Majorana pairs $i \gamma_{2i} \gamma_{2i+1}$. 
%It means that the ensemble average of the module of the expectation value $\langle{Z_{m}Z_{n}}\rangle$ is always finite as long as the pairing configuration represented by the red double solid lines is favored in the dynamics. 
%Similar observations can also be made for the other two order parameters. 
%This explains why the modified order parameters can be used to detect the corresponding long-range orders stabilized in the circuit dynamics. 

% In fact, the unmeasured two Majorana modes can also be detected by the purification dynamics of the whole system from an initial maximally mixed state. 
% At the initial time $t=0$, no Majorana modes is paired and the system entropy is $S(t=0) = L$. 
% As the time increases, more and more Majorana modes are paired with each other and $S(t)$ will decrease in the circuit dynamics. 
% The value of $S(t)$ is just $L$ minus the pairing number of the Majorana modes at time $t$. 
% As a result, the unmeasured two Majorana modes give rise to $S(t \gg L) = 1$ in the purification dynamics.

To end this part, we would like to mention that the Majorana modes can be split into two sets, say, $\mathcal{S}_{1} = \{ \gamma_{2i-1} \}_{i=1}^{L}$ and $\mathcal{S}_{2} = \{ \gamma_{2i} \}_{i=1}^{L}$, such that any projective measurement in Eq.~\eqref{eq:ci_measurement} contains one Majorana mode from $\mathcal{S}_{1}$ and the other from $\mathcal{S}_{2}$.
Therefore, the wordline orientability of the corresponding loop model is conserved~\cite{klocke2023prx}. 
As a result, there is no critical ``Goldstone'' phase in our case, in contrast to the completely packed loop model with crossings where the orientability symmetry is broken~\cite{nahum2013prb}.

\bibliographystyle{apsrev4-2}
% \bibliography{ref}
\let\oldaddcontentsline\addcontentsline% Store \addcontentsline
\renewcommand{\addcontentsline}[3]{}% Make \addcontentsline a no-op
\bibliography{main_v2.bib}

\let\addcontentsline\oldaddcontentsline% Restore \addcontentsline
\onecolumngrid

\clearpage
\newpage

%%%%%%%%%%%%%%%%%%%%%% SM %%%%%%%%%%%%%%%%%%%%%%
\widetext

\begin{center}
\textbf{\large Supplemental Material for ``Gapless Symmetry-Protected Topological States in Measurement-Only Circuits''}
\end{center}

\maketitle

\renewcommand{\thefigure}{S\arabic{figure}}
\setcounter{figure}{0}
\renewcommand{\theequation}{S\arabic{equation}}
\setcounter{equation}{0}
\renewcommand{\thesection}{\Roman{section}}
\setcounter{section}{0}
\setcounter{secnumdepth}{4}

\addtocontents{toc}{\protect\setcounter{tocdepth}{0}}
{
\tableofcontents
}

\section{Observables in measurement-only quantum circuits}
\label{sm:quantity}

In this section, we introduce the physical observables utilized in the main text to investigate the steady state of measurement-only circuits.

\emph{Half-chain entanglement entropy.} For the state $|\psi_{t}\rangle$ evolved after $t$ discrete time steps, we can calculate the von Neumann entanglement entropy, 
% $S_{R}$, 
\begin{equation}
    S_{R} = - \text{Tr}(\rho_{R} \log_{2}\rho_{R}) \,,
\end{equation}
where $\rho_{R}$ is the reduced density matrix of the subregion $R$ of the system and $S_{R}$ quantifies the entanglement between the subsystem $R$ and its complement $\bar{R}$.
Then the most common observable in measurement-only circuits, the half-chain entanglement entropy, is defined by $S_{\rm Half} \equiv S_{\mathcal{L}}$ with $\mathcal{L} = \{1,\cdots L/2\}$, whose size-scaling behaviors can be used to classify the steady states into area-law or volume-law entangled phases and to detect the entanglement phase transitions. 
In particular, for the percolation-type phase transition relevant in measurement-only circuits, it has been investigated that the ensemble averaged $S_{\text{Half}}$ at the criticality exhibits a logarithmic scaling with the system size $L$~\cite{klocke2023prx}, namely, $S_{\rm Half}(L) = c_\text{eff} \log_{2}\!L + c'$, where the prefactor $c_\text{eff}$ which we call the effective central charge has an exact value $\sqrt{3}\ln2/(4\pi)$ under open boundary conditions and $c'$ is a non-universal constant. 

\emph{Generalized topological entropy.} Another entanglement quantity relevant in our work is the generalized topological entanglement entropy $S_{\text{topo}}$ which can be used to characterize the nontrivial topology of the evolved state $|\psi_{t}\rangle$. 
By partitioning the whole system into four subregions with equal size, $A|B|D|C$ [see Fig.~2(a) in the main text], $S_{\rm topo}$ is defined by~\cite{wen2019book}
\begin{equation}
    \label{eq:entropy_topo}
    S_{\text{topo}} \equiv S_{AB}+S_{BC}-S_{B}-S_{ABC} \,.
\end{equation}
For example, it is known that $S_{\rm topo} = 2$ within the $\mathbb{Z}_{2} \times \mathbb{Z}_{2}$ (cluster) symmetry-protected topological (SPT) phase while $S_{\rm topo} = 0$ in the spontaneous symmetry-breaking (SSB) phase. 
Therefore, $S_{\rm topo}$ can faithfully distinguish between the SPT and SSB phases, as well as locate the corresponding critical point~\cite{lavasani2021measurement}. 

% mutual information is not used in the main text, so I move it to Sec.IIID

\emph{Order parameters.} Besides the entanglement observables enumerated above, some recent works~\cite{sang2021prr,morral2023prb} show that conventional order parameters can also be used to characterize the long-range orders stabilized by the quantum circuit with suitable modifications. 
For the Ising cluster circuit model, we can define three order parameters to detect the possible existence of the paramagnetic (PM), SSB, and SPT orders
\begin{align}
    \label{eq:orders}
    O_{\rm SSB}(i,j) & = \overline{|\langle Z_{i} Z_{j} \rangle|} \,, \\
    O_{\rm PM}(i,j) & = \overline{|\langle X_{i} X_{i+1} \cdots X_{j-1} X_{j} \rangle|} \,, \\
    O_{\rm SPT}(i,j) & = \overline{|\langle Z_{i-1} Y_{i} X_{i+1} \cdots X_{j-1} Y_{j} Z_{j+1} \rangle|} \,,
\end{align}
where, in practical simulations, one can choose $i=L/4+1$ and $j=3L/4$ to reduce the possible boundary effect under the open boundary condition.
Different from the conventional definitions used in the equilibrium case, here, the observables are ensemble averaged over many different trajectory realizations and the module before the average is necessary to obtain meaningful (nonzero) results~\cite{morral2023prb}. 
For the $\mathbb{Z}_{4}$ symmetric circuit model, similar order parameters can be defined as shown in Section~III of the Supplemental Material.

\section{Additional results for the Ising cluster circuit model}
\label{sm:z2}

\subsection{Steady state phase diagram via the order parameters}

In the main text, the phase transition between the SPT and SSB phases has been investigated mainly in the case of no $X_{i}$ measurement, namely, $p_{\rm x} = 0$.
Here, we aim to consider the case where all three types of measurements are present and try to map out the phase diagram of the steady state in the whole parameter space $(p_{\rm x}, p_{\rm zz}, p_{\rm zxz})$. 
It is noted that previous works have studied the cases of $p_{\rm zxz} = 0$~\cite{lavasani2021measurement,sang2021prr} and $p_{\rm zz} = 0$~\cite{morral2023prb}. 
It was observed that there is a phase transition between the SSB and trivial PM phases in the former and a transition between the SPT and PM phases in the latter; both transitions happen when the two relevant measurement probabilities are tuned to be equal. 
However, when all three types of measurements are present in the circuit, the situation becomes more complex and the phase diagram has not been figured out before.

\begin{figure}
    \centering
    \includegraphics[width=0.85\linewidth]{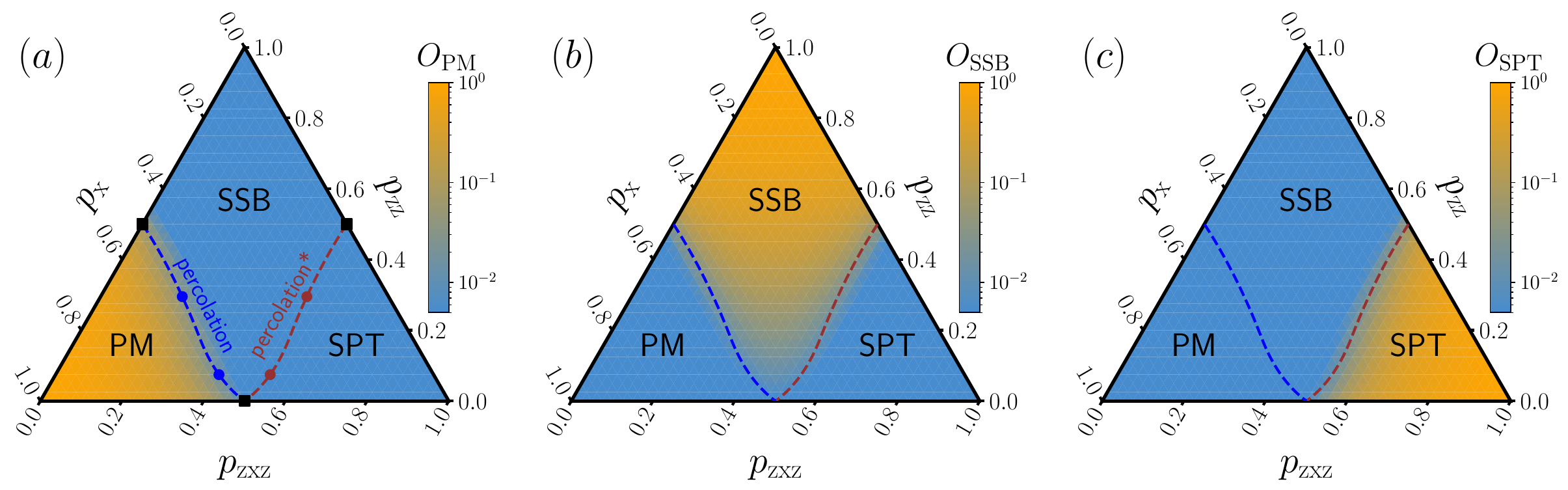}
    \caption{Order parameters, (a) $O_{\rm PM}$, (b) $O_{\rm SSB}$, and (c) $O_{\rm SPT}$, shown as a function of the measurement probabilities $p_{\rm x}$, $p_{\rm zz}$, and $p_{\rm zxz}$, simulated with system size $L = 96$ under open boundary conditions. The black squares are exact critical points when only two types of measurements are present. The red circles are critical points obtained by the data collapse of the generalized topological entanglement entropy (see Fig.~\ref{fig:sm_z2_Stopo}) and the red dashed line is a guide to the eye. The blue circles and the corresponding dashed line are obtained by $p_{\rm x} \leftrightarrow p_{\rm zxz}$.}
    \label{fig:order_diagram}
\end{figure}

To map out the steady-state phase diagram, we first calculate the three order parameters, $O_{\text{PM}}$, $O_{\text{SSB}}$, and $O_{\text{SPT}}$, within the whole parameter regime as displayed in Fig.~\ref{fig:order_diagram}. 
It is obvious that the phase diagram is roughly partitioned into three regions: SSB, PM, and SPT.
The result supports the stability of these three phases in the presence of all three competing measurements.

\subsection{Determination of the critical points for the case of $p_\text{x} = 0.2$ and $p_\text{x} = 0.4$}

Our next task is to determine the location of the transition lines between the different phases. 
Similar to the duality argument given in Ref.~\cite{lavasani2021measurement}, it is noted that the structure of the phase diagram should be symmetric for $p_{\rm x} \leftrightarrow p_{\rm zxz}$ under the global unitary transformation $U_{\rm SPT} = \prod_{i=1}^{L} \text{CZ}_{i,i+1}$ (here $\text{CZ}_{i,i+1}$ is the two-qubit controlled-Z gate) which transforms $X_{i}$ into $Z_{i-1}X_{i}Z_{i+1}$ and vice versa; periodic boundary condition is assumed here. 
Since $U_{\rm SPT}$ transforms local stabilizers to local stabilizers, the state in an area-law entangled phase still obeys the area law after the transformation~\cite{lavasani2021measurement}. 
As a result, if a continuous phase transition occurs at $(p_{\rm x},p_{\rm zz},p_{\rm zxz})$ with logarithmic entanglement entropy, we can find another corresponding phase transition intermediately by $p_{\rm x} \leftrightarrow p_{\rm zxz}$. 
We also notice that this symmetry becomes transparent in the Majorana representation as illustrated in Fig.~5 in the Appendix.
Therefore, it is sufficient to locate the transition line between SPT and SSB, and the other one can be inferred simply by $p_{\rm x} \leftrightarrow p_{\rm zxz}$. 

\begin{figure}
    \centering
    \includegraphics[width=0.75\linewidth]{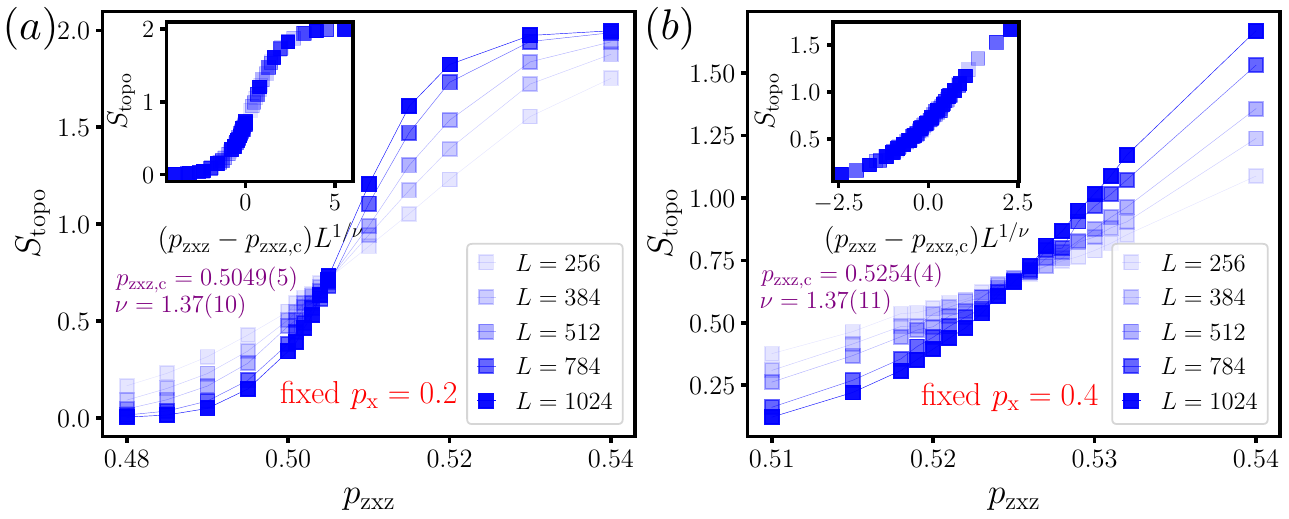}
    \caption{The generalized topological entanglement entropy $S_{\rm{topo}}$ versus the measurement probability $p_{\text{zxz}}$ for various system sizes with (a) $p_\text{x} = 0.2$ and (b) $p_\text{x} = 0.4$, respectively. The insets exhibit the data collapses of $S_{\text{topo}}$ with $p_{\text{zxz},c}=0.5049(5)$ and $\nu = 1.37(10)$ for $p_\text{x} = 0.2$ and $p_{\text{zxz},c}=0.5254(4)$ and $\nu = 1.37(11)$ for $p_\text{x} = 0.4$, respectively.}
    \label{fig:sm_z2_Stopo}
\end{figure}

In the main text, we have evidenced that the generalized topological entanglement entropy $S_{\rm topo}$ is a powerful tool to study the SPT-SSB transition. 
By fixing $p_{\rm x} = 0.2$, we have computed the ensemble average of $S_{\rm topo}$ as a function of $p_{\rm zxz}$ as exhibited in Fig.~\ref{fig:sm_z2_Stopo}(a). 
It is shown that curves of different system sizes cross at a single point suggesting the existence of a phase transition.
The data collapse further determines the transition point $p_{\text{zxz},c} = 0.5049(5)$ and the critical exponent $\nu = 1.37(10)$. 
Similar analysis for the case of $p_\text{x} = 0.4$ also gives the estimation $p_{\text{zxz},c} = 0.5254(4)$ and $\nu = 1.37(11)$\,. 
Together with the logarithmic half-chain entanglement entropy displayed in Fig.~2(b) in the main text, it implies that adding a finite probability of $X_{i}$ measurement does not change the bond percolation universality class of the SPT-SSB transition.
% which is the bond percolation.

\subsection{Purification dynamics on the symmetry-enriched percolation line for $p_\text{x} \neq 0$}

\begin{figure}
    \centering
    \includegraphics[width=0.65\linewidth]{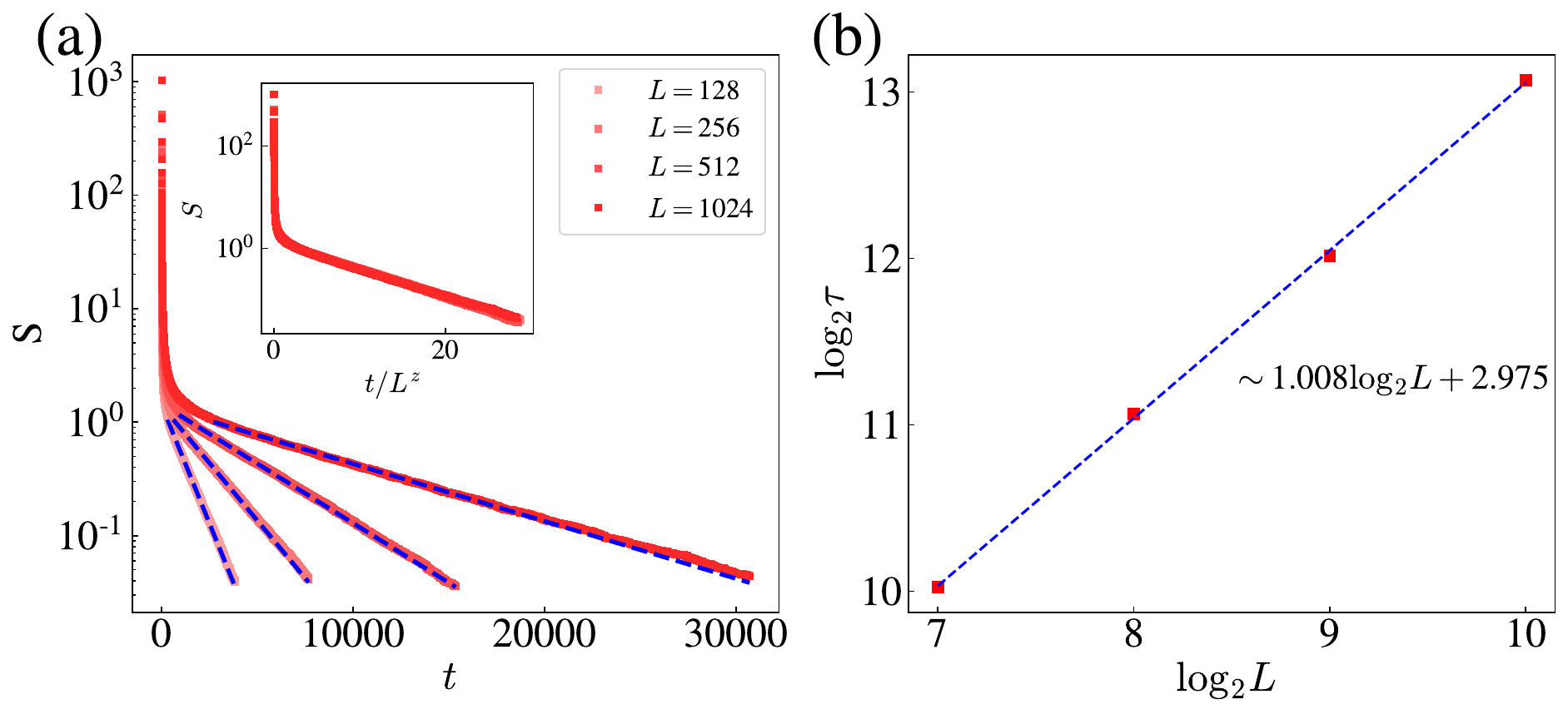}
    \caption{(a) The entanglement entropy of the whole system $S$ as a function of the evolving time $t$ for various system sizes at the estimated critical point $p_{\rm{x}}=0.2$ and $p_{\rm{zxz}}=0.5049$\,. 
    The initial state is the maximally mixed state. 
    The dashed blue line represents the fitting $S(t) \sim e^{-t/\tau}$ using the late-time data. Inset shows the data collapse of $S(t)$ with rescaled x-axis $t/L^{z}$ where $z$ is given by $\tau \sim L^{z}$. 
    (b) shows the least-squares fitting of $\tau \sim L^{z}$, which gives $z \approx 1.008$\,.}
    \label{fig:sm_z2_px0.2_S}
\end{figure}

\begin{figure}
    \centering
    \includegraphics[width=0.65\linewidth]{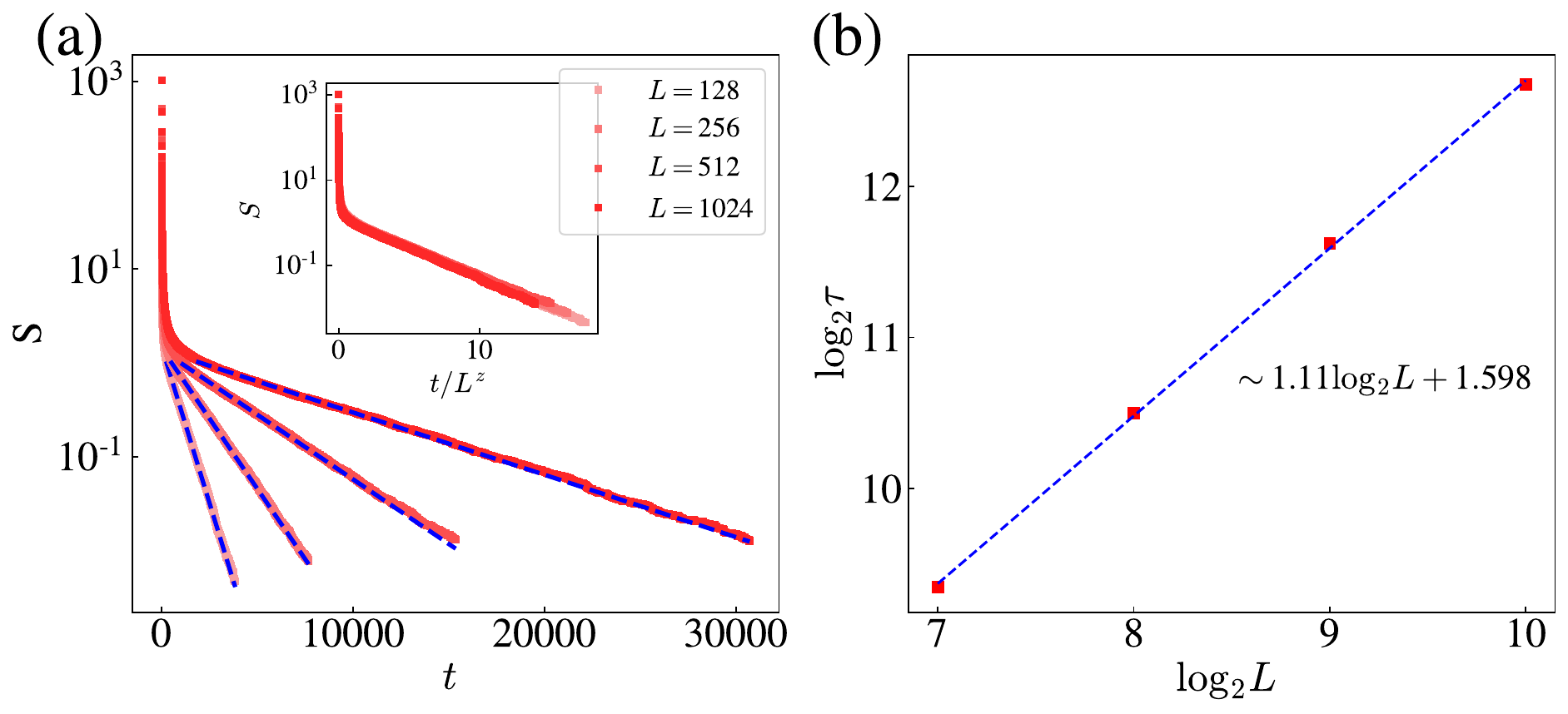}
    \caption{(a) The dynamics of the entanglement entropy of the whole system $S$ starting from the maximally mixed state for different system sizes at the estimated critical point $p_{\rm{x}}=0.4$ and $p_{\rm{zxz}}=0.5254$\,. 
    The dashed blue line represents the fitting $S(t) \sim e^{-t/\tau}$ using the late-time part of the data. Inset shows the data collapse with rescaled x-axis $t/L^{z}$ where $z$ is given by $\tau \sim L^{z}$. 
    (b) displays the least-squares fitting $\tau \sim L^{z}$, which gives $z \approx 1.11$\,.}
    \label{fig:sm_z2_px0.4_S}
\end{figure}

In this section, we performed additional numerical simulations for the purification dynamics at the symmetry-enrich percolation critical points in the case of $p_\text{x} = 0.2$ and $p_\text{x} = 0.4$\,. 
By starting from a maximally mixed state, the initial entanglement entropy of the full system is $S(t=0) = L$ and the entanglement entropy will decrease as time increases as the projective measurements can purify the state. 
As presented in Fig.~\ref{fig:sm_z2_px0.2_S}, in the late-time regime, we can fit the residue entropy $S(t) \sim e^{-t/\tau}$ and the fitted $\tau$ is found satisfying $\tau \sim L^{z}$ where $z=1.008$ is the dynamical exponent. 
The data collapse in the inset indicates that the residue entropy of the full system is always nonzero at finite times in the thermodynamic limit.
It means that there is an encoded subspace that survives for arbitrarily long times when $L\to\infty$, providing potential evidence for the existence of the edge modes on the symmetry-enriched percolation line.
Numerical simulations for the case of $p_\text{x} = 0.4$ have also been performed and similar results are obtained as displayed in Figs.~\ref{fig:sm_z2_px0.4_S}.

\subsection{Different behaviors of the string operators at the topologically trivial and nontrivial critical points}

\begin{figure}
    \centering
    \includegraphics[width=0.65\linewidth]{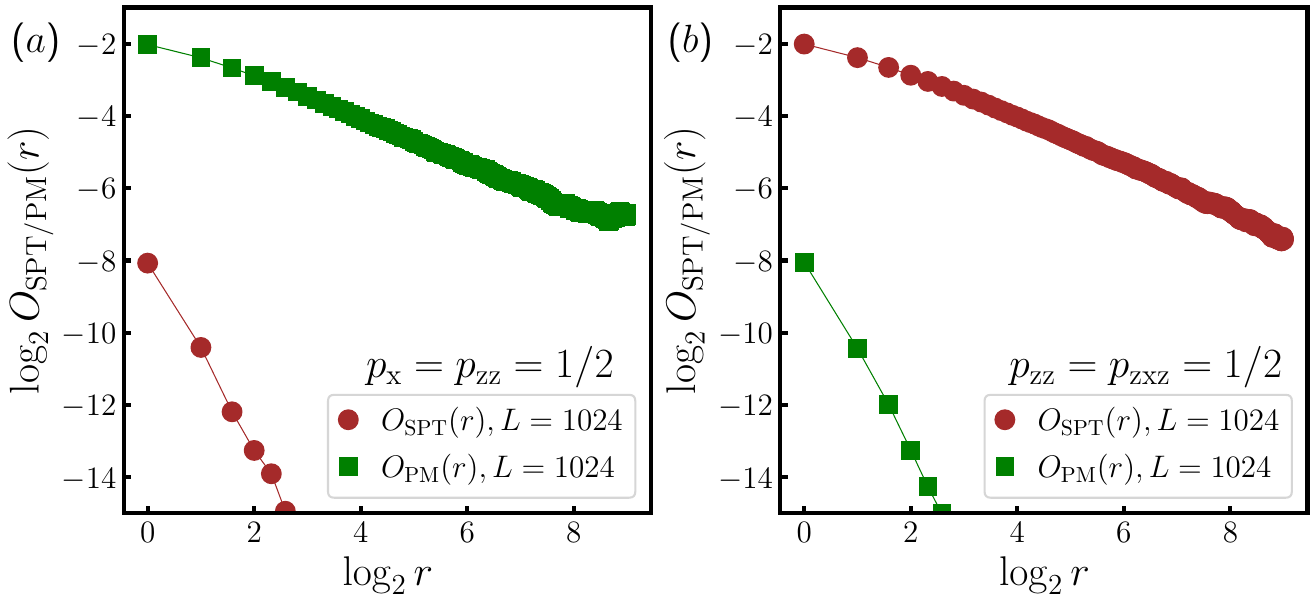}
    \caption{String order parameters, $O_{\text{SPT}}(r)$ (red circles) and $O_{\text{PM}}(r)$ (green squares), as a function of the site distance $r$ for (a) $p_{\text{x}} = p_{\text{zz}} = 1/2$ and (b) $p_{\text{zz}} = p_{\text{zxz}} = 1/2$, respectively. The simulated system size is $L = 1024$ under open boundary conditions.}
    \label{fig:ci_compare}
\end{figure}

Finally, we compare the scaling behaviors of the relevant string order parameters $O_{\text{SPT}}(r)$ and $O_{\text{PM}}(r)$ at the critical points $p_{\text{x}} = p_{\text{zz}} = 1/2$ and $p_{\text{zz}} = p_{\text{zxz}} = 1/2$, respectively. 
As shown in Fig.~\ref{fig:ci_compare}, $O_{\text{SPT}}(r)$ exhibits a power-law decay with respect to the site distance $r$ at the symmetry-enriched percolation critical point, while $O_{\text{PM}}(r)$ decays much faster than $O_{\text{SPT}}(r)$ at this point. 
However, at the trivial percolation criticality, the situation is reversed, and $O_{\text{SPT}}(r)$ decays significantly faster than $O_{\text{PM}}(r)$. 
This result provides further evidence for the topological distinction between the two percolation critical lines shown in Fig.~1(b) in the main text.

\section{Additional results for the $\mathbb{Z}_{4}$ symmetric circuit model}
\label{sm:z4}

\subsection{The equilibrium ground-state phase diagram}

\begin{figure}[h]
    \centering
    \includegraphics[width=0.4\linewidth]{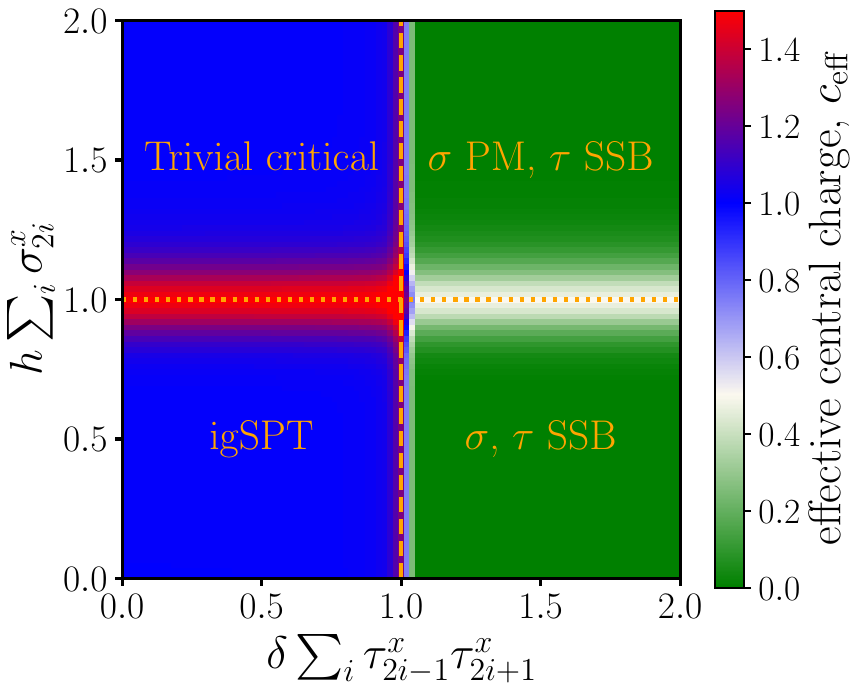}
    \caption{The ground-state phase diagram of the Hamiltonian, $H_\text{igSPT} - \delta \sum_{i} \tau_{2i-1}^{x} \tau_{2i+1}^{x} - h \sum_{i} \sigma_{2i}^{x}$, as a function of $h$ and $\delta$ is mapped out by calculating the effective central charge $c_\text{eff}$, where $c_\text{eff}$ is estimated by fitting to $S(l) = \frac{c_\text{eff}}{3} \log_{2} (L/\pi \sin[\pi l/L]) + \text{const}$. The dotted line ($h = 1$) is an Ising transition while the dashed line ($\delta = 1$) is a BKT transition. Simulation is performed by exploiting the density matrix renormalization group algorithm~\cite{white1992prl,white1993prb,SCHOLLWOCK201196} with $L = 24$ under periodic boundary conditions.}
    \label{fig:sm_z4_gs_diagram}
\end{figure}

First, we recall the physics in the equilibrium case for later comparison with its non-equilibrium counterpart.
The model is a quantum spin chain with two $1/2$ spins per unit cell, which is described by the following Hamiltonian, 
\begin{equation}
    \label{eq:gSPT}
    H_{\rm igSPT} = - \sum_{i=1}^{L} \big( \tau_{2i-1}^{z}\sigma_{2i}^{x}\tau_{2i+1}^{z} +  \tau_{2i-1}^{y}\sigma_{2i}^{x}\tau_{2i+1}^{y} + \sigma_{2i}^{z}\tau_{2i+1}^{x}\sigma_{2i+2}^{z} \big) \,.
\end{equation}
This model is obtained by stacking an Ising ($\mathbb{Z}_{2}$ SSB) spin chain with an XX Hamiltonian,
\begin{equation}
    \label{eq:Ising&XX}
    H_{\rm Ising + XX} = - \sum_{i=1}^{L} \big( \tau_{2i-1}^{z}\tau_{2i+1}^{z} + \tau_{2i-1}^{y}\tau_{2i+1}^{y} \big) - \sum_{i=1}^{L} \sigma_{2i}^{z}\sigma_{2i+2}^{z} \,,
\end{equation}
through the Kennedy-Tasaki (KT) transformation~\cite{li2023intrinsicallypurelygaplesssptnoninvertibleduality}.
Since the Ising chain and the XX chain is completely decoupled in $H_{\rm Ising+XX}$, we can easily read off the spin-spin correlation functions
\begin{equation}
    \langle \sigma_{2i}^{z}\sigma_{2j}^{z} \rangle = 1, \quad{} \langle \tau_{2i-1}^{z}\tau_{2j-1}^{z} \rangle \sim 1 / |i-j|^{\Delta} \,,
\end{equation}
where $\Delta/2$ is the scaling dimension of $\tau^{z}$. 
These conventional correlations then become the string order parameters after the KT transformation
\begin{equation}
    O_{\rm \sigma}(r = |i-j|) = \langle \sigma_{2i}^{z} (\prod_{k=i}^{j-1} \tau_{2k+1}^{x}) \sigma_{2j}^{z} \rangle = 1, \quad O_{\rm \tau}(r = |i-j|) = \langle \tau_{2i-1}^{z} (\prod_{k=i}^{j-1} \sigma_{2k}^{x}) \tau_{2j-1}^{z} \rangle \sim 1 / |i-j|^{\Delta} \,.
\end{equation}
These nonlocal string operators characterize the nontrivial topology of the intrinsic gSPT phase. 
More specifically, the system possesses a $\mathbb{Z}_{4}$ symmetry generated by $U=\prod_{i}\sigma_{2i}^{x} e^{i\frac{\pi}{4}(1-\tau_{2i-1}^{x})}$, which exhibits an emergent anomaly at low energies that is the same anomaly on the boundary of a 2+1D Levin-Gu SPT state~\cite{levin2012prb}. 
One can see easily that, under the open boundary condition, the square of $U$ fractionalizes onto each end of the boundary as $U^{2} \sim \tau_1^{x} \sigma_2^{z} \sigma_{2L}^{z}$~\cite{wen2023classification11dgaplesssymmetry,wen2023prb}. 
It is also noted that the gSPT phase is robust against the symmetric perturbation by adding $\delta \sum_{i} \tau_{2i-1}^{x} \tau_{2i+1}^{x}$ when $|\delta| < 1$ or $h \sum_{i} \sigma_{2i}^{x}$ when $|h| < 1$; 
the ground-state phase diagram in the presence of both perturbations is mapped out in Fig.~\ref{fig:sm_z4_gs_diagram}. 
It motivates us to investigate the steady state phase diagram of the $\mathbb{Z}_{4}$ symmetric measurement-only circuit including the corresponding two-site and one-site projective measurement perturbations. 
The resulting non-equilibrium phase diagram is displayed in Fig.~3(a1) in the main text.

\subsection{Simulation results for the case of $p_{h} = p_{\delta}$}

In the main text, we have investigated the physics of the steady state of the $\mathbb{Z}_{4}$ symmetric circuit model along the horizontal $p_{h} = 0.0$ line and the vertical $h_{\delta} = 0.08$ line. 
Here, we provide additional results for the case of $p_{h} = p_{\delta}$ which includes the crossing point of the BKT and percolation transition lines, namely, $p_{h} = p_{\delta} = 1/5$. 

\begin{figure}
    \centering
    \includegraphics[width=\linewidth]{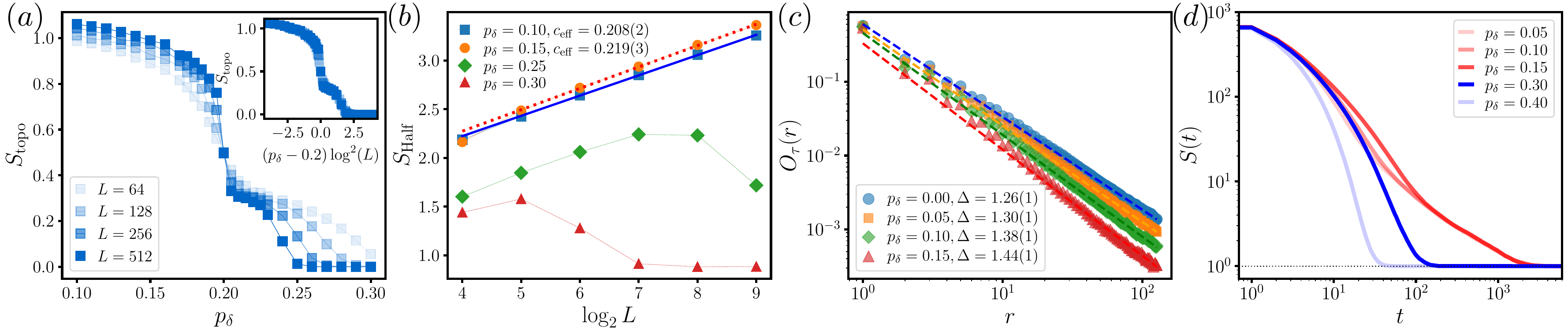}
    \caption{The simulation result for the $\mathbb{Z}_{4}$ symmetric circuit model in the case of $p_{h} = p_{\delta}$ crossing the special point $p_{h} = p_{\delta} = 1/5$. (a) The data collapse of the generalized topological entanglement entropy $S_\text{topo}$ with the rescaled x-axis $(p_{\delta} - 0.2) \log^{2}(L)$ using the critical point $p_{\delta,c} = 1/5$. (b) The finite-size scaling of the half-chain entanglement entropy $S_{\text{Half}}$ shows logarithmic growth for $p_{\delta} < 1/5$. The blue solid and orange dotted lines are least-squares fittings giving the estimated effective central charges $c_{\rm eff} = 0.208(2)$ and $c_{\rm eff} = 0.219(3)$, respectively. (c) The string order parameter $O_{\tau}(r)$ as a function of the site distance $r$ for different $p_{\delta} < 1/5$. (d) The dynamics of the full system entropy $S(t)$ versus the evolution time $t$ for different values of $p_{\delta}$. The simulated system size is $L = 256$ for (c) and $L=512$ for (d) under open boundary conditions.}
    \label{fig:z4_ph_pdelta}
\end{figure}

As displayed in Fig.~\ref{fig:z4_ph_pdelta}(a), we calculate the generalized topological entanglement entropy $S_\text{topo}$ for different system sizes near the critical point; 
a perfect data collapse is achieved by using $p_{\delta,c} = 1/5$ with a standard logarithmic-squared scaling form for BKT transitions. 
It is noted that the coexisting percolation transition is invisible in the finite-size scaling analysis due to the BKT transition.
By further investigating the system-size dependence of the half-chain entanglement entropy $S_{\text{Half}}$, we can clearly see that $S_{\text{Half}}$ shows a logarithmic growth with $L$ at $p_{\delta} = 0.10$ and $p_{\delta} = 0.15$ (belongs to the gSPT phase) as shown in Fig.~\ref{fig:z4_ph_pdelta}(b). 
The fitted effective central charge $c_\text{eff}$ is also close to $2 \times \frac{\sqrt{3}\ln2}{4\pi}$. 
The parameter regime $p_{\delta} > 1/5$ is an (a) SSB (PM) in the $\tau$ ($\sigma$) degrees of freedom, which shows an area-law $S_{\text{Half}}$ when $L$ is large enough. 
This observation is also supported by examining the asymptotic behavior of the string order parameter $O_{\tau}(r)$ as plotted in Fig.~\ref{fig:z4_ph_pdelta}(c). 
It is clear that $O_{\tau}(r)$ exhibits a power-law decaying behavior with an exponent $\Delta \approx 4/3$ for $p_{\delta} < 1/5$ supporting the existence of a stable gSPT phase. 
Finally, we can see a nonzero residue entropy $S(t \gg L) = 1$ in the purification dynamics for both $p_{\delta} < 1/5$ and $p_{\delta} > 1/5$ as shown in Fig.~\ref{fig:z4_ph_pdelta}(d). 
The nontrivial residue entropy comes from the topological protected edge modes in the gSPT phase for $p_{\delta} < 1/5$ and the SSB in the $\tau$ degrees of freedom for $p_{\delta} > 1/5$, respectively.

In summary, the numerical results shown here agree with the observation in the main text and can help us to check the correctness of the predicted BKT (percolation) line $p_{h} + 4p_{\delta} = 1$ ($4p_{h} + p_{\delta} = 1$) in the steady-state phase diagram shown in Fig.~3(a1) in the main text.

\subsection{Simulation results for spin-spin correlations}

\begin{figure}
    \centering
    \includegraphics[width=0.85\linewidth]{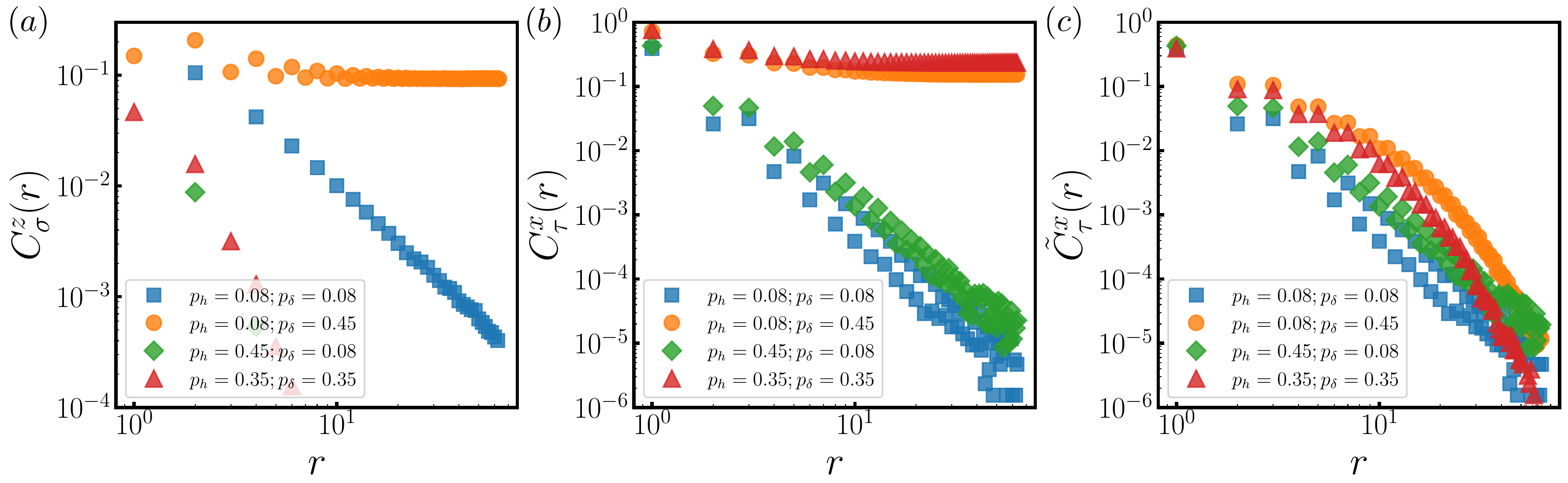}
    \caption{Spin-spin correlations (a) $C_{\sigma}^{z}(r)$ and (b) $C_{\tau}^{x}(r)$ versus the site distance $r$ for four representative parameter points in the steady-state phase diagram. (c) The connected correlation $\tilde{C}_{\tau}^{x}(r)$ as a function of $r$ for the same parameters. Simulations are performed with a system of size $L = 128$ under open boundary conditions.}
    \label{fig:sm_z4_correlation}
\end{figure}

To support the analyses and the conclusions made in the main text, we perform simulations to calculate the spin-spin correlations respectively for $\sigma$ and $\tau$ degrees of freedom. 
In particular, we investigate the following (connected) correlations
\begin{equation}
    C_{\sigma}^{z}(r) = \overline{ | \langle{\sigma_{2}^{z}\sigma_{2+2r}^{z}}\rangle | } \,, \qquad{} C_{\tau}^{x}(r) = \overline{ | \langle{\tau_{1}^{x}\tau_{1+2r}^{x}}\rangle | } \,, \qquad{} \tilde{C}_{\tau}^{x}(r) = \overline{ | \langle{\tau_{1}^{x}\tau_{1+2r}^{x}}\rangle - \langle{\tau_{1}^{x}}\rangle \langle{\tau_{1+2r}^{x}}\rangle | } \,,
\end{equation}
at representative points for each phase shown in Fig.~3(a1) in the main text. 
We do not consider the connected $\sigma^{z}$ correlation as the expectation $\langle{\sigma_{i}^{z}}\rangle$ is zero for all $i$ which is guaranteed by the symmetry of the circuit dynamics. 

As displayed in Fig.~\ref{fig:sm_z4_correlation}, the spin-spin correlations show different behaviors among the chosen points. 
For $p_{h} = p_{\delta} = 0.08$ in the Phase~I, both $C_{\sigma}^{z}(r)$ and $C_{\tau}^{x}(r)$ display power-law decaying behaviors due to the critical $\tau$ degrees of freedom. 
When we add the one-site perturbation and drive the system into the Phase~II (e.g., $p_{h} = 0.45$ and $p_{\delta} = 0.08$), the $\tau$ degrees of freedom are still critical while the $\sigma$ degrees of freedom are trivially gapped leading to an exponentially decaying $C_{\sigma}^{z}(r)$. 
On the other hand, when we add the two-site perturbation and drive the system into the Phase~III (e.g., $p_{h} = 0.08$ and $p_{\delta} = 0.45$), the $\tau$ degrees of freedom are gapped out resulting in a long-range SSB order in the $x$ ($z$) direction of the $\tau$ ($\sigma$) spins. 
Finally, for $p_{h} = p_{\delta} = 0.45$ in the Phase~IV, the system exhibits PM (SSB) in the $\sigma$ ($\tau$) degrees of freedom and we have an (a) exponential decaying (long-range) $C_{\sigma}^{x}$ ($C_{\tau}^{x}$). 

To end this part, we notice that the exponential decay of the connected correlation $\tilde{C}_{\tau}^{x}(r)$ in Phase~III and Phase~IV means that the $\tau$ degrees of freedom are stabilized into a symmetry-breaking state instead of a GHZ state. 
The $\tau$ degrees of freedom do not evolve to a GHZ state due to the specific choice of the initial state in our work.

\subsection{Effect of the single-site measurement $\tau_{2i-1}^{x}$ on the $\mathbb{Z}_{4}$ symmetric circuit model}

\begin{figure}
    \centering
\includegraphics[width=0.6\linewidth]{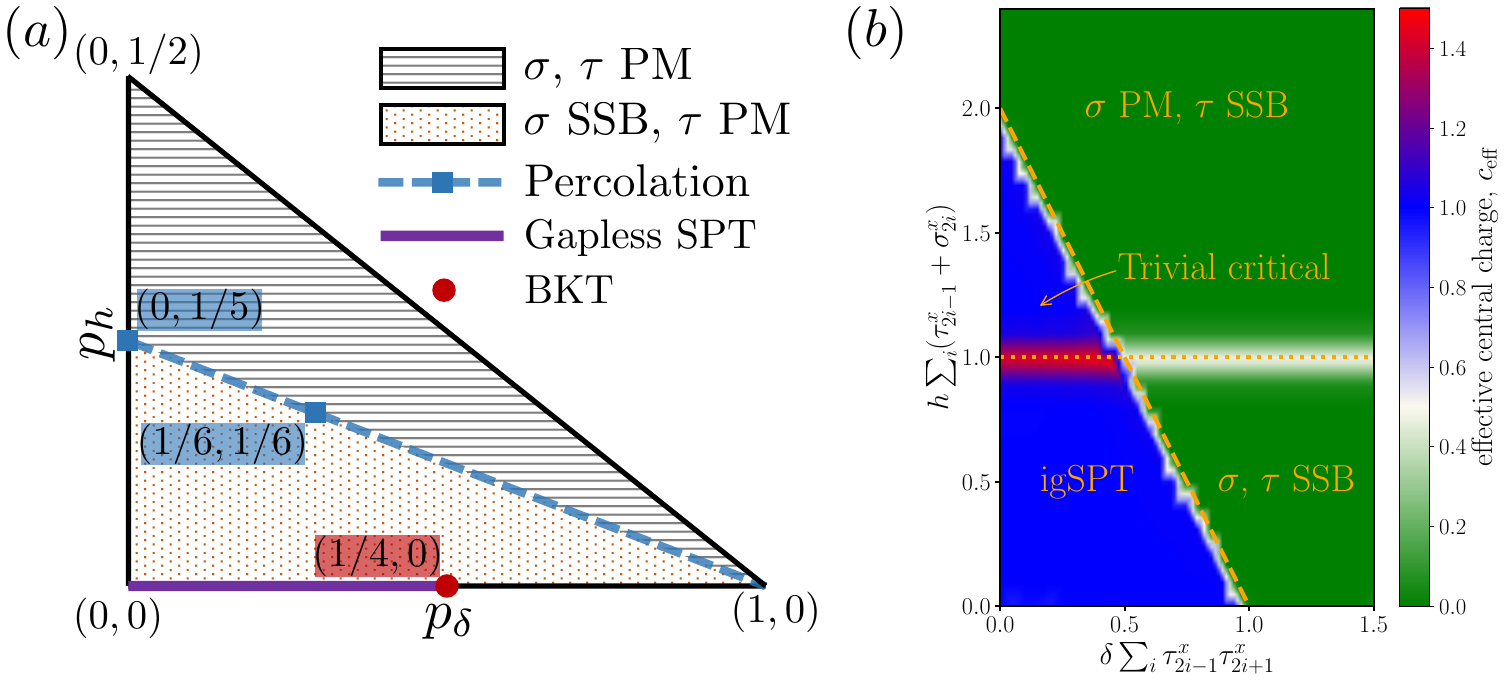}
    \caption{(a) The steady state phase diagram of the $\mathbb{Z}_{4}$ symmetric circuit including the single-site perturbation $\tau_{2i-1}^{x}$ with the same occurring probability as $\sigma_{2i}^{x}$; it is noted that $3p_{t}+p_{\delta}+2p_{h}=1$. The percolation transition line (blue dashed) is $5p_{h} + p_{\delta} = 1$.
    (b) The corresponding ground-state phase diagram of the Hamiltonian, $H_\text{igSPT} - \delta \sum_{i} \tau_{2i-1}^{x} \tau_{2i+1}^{x} - h \sum_{i} ( \tau_{2i-1}^{x} + \sigma_{2i}^{x} )$, is obtained from calculating the effective central charge $c_\text{eff}$. The dotted line ($h = 1$) is an Ising transition while the dashed line ($2\delta + h = 2$) is a BKT transition. Simulation is performed by exploiting the density matrix renormalization group algorithm with $L = 48$ under periodic boundary conditions.}
    \label{fig:sm_z4v2_diagram}
\end{figure}

In this section, we investigate the effect of the single-site operator $\tau_{2i-1}^{x}$ on the $\mathbb{Z}_{4}$ symmetric circuit model studied in the main text. 
Specifically, we set the occurring probability of the $\tau_{2i-1}^{x}$ measurement same as $\sigma_{2i}^{x}$ and the probabilities satisfy $3p_{t} + p_{\delta} + 2p_{h} = 1$. 
As mentioned in the main text, the single-site measurement $\tau_{2i-1}^{x}$ is a relevant perturbation at the fixed point with two copies of percolation (e.g., the point $p_{h} = p_{\delta}$ = 0)~\cite{klocke2023prx}. 
Therefore, the inclusion of the $\tau_{2i-1}^{x}$ measurement results in the disappearance of a stable gSPT region, leading to a different steady-state phase diagram, as shown in Fig.~\ref{fig:sm_z4v2_diagram}(a). 
This is in stark contrast to its equilibrium counterpart where $h \sum_{i} \tau_{2i-1}^{x}$ is an irrelevant perturbation, as depicted in Fig.~\ref{fig:sm_z4v2_diagram}(b). 
Since the line $p_{h} = 0$ has been investigated in the phase diagram shown in Fig.~3(a1) in the main text, here, we focus on the case of $p_{h} > 0$ and study the two non-topological area-law phases therein.

\begin{figure}
    \centering
    \includegraphics[width=\linewidth]{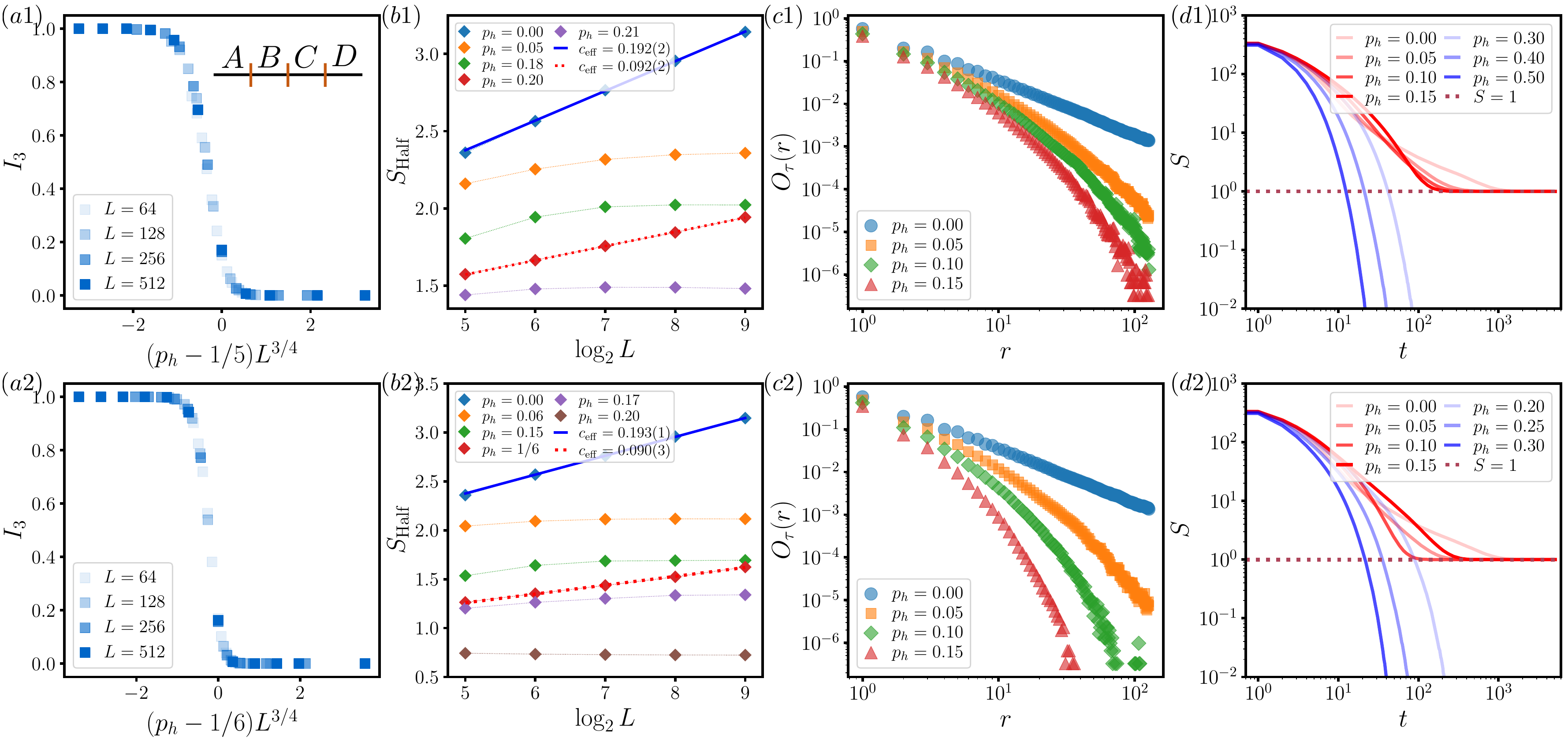}
    \caption{The simulation results for the $\mathbb{Z}_{4}$ symmetric circuit model including the single-site perturbation $\tau_{2i-1}^{x}$ respectively in the case of $p_{\delta} = 0$ [(a1)-(d1)] and $p_{h} = p_{\delta}$ crossing the special point $p_{h} = p_{\delta} = 1/6$ [(a2)-(d2)]. (a1-a2) The data collapse of the generalized topological entanglement entropy $S_\text{topo}$. (b1-b2) The finite-size scaling of the half-chain entanglement entropy $S_{\text{Half}}$ shows logarithmic growth at the percolation critical points. The blue solid and red dotted lines are least-squares fittings. (c1-c2) The string order parameter $O_{\tau}(r)$ as a function of the site distance $r$. (d1-d2) The purification dynamics of the full system entropy $S(t)$ versus the evolution time $t$. The simulated system size is $L = 256$ for (c1-c2) and $L=512$ for (d1-d2) under open boundary conditions.}
    \label{fig:sm_z4v2_main}
\end{figure}

\begin{figure}
    \centering
    \includegraphics[width=0.75\linewidth]{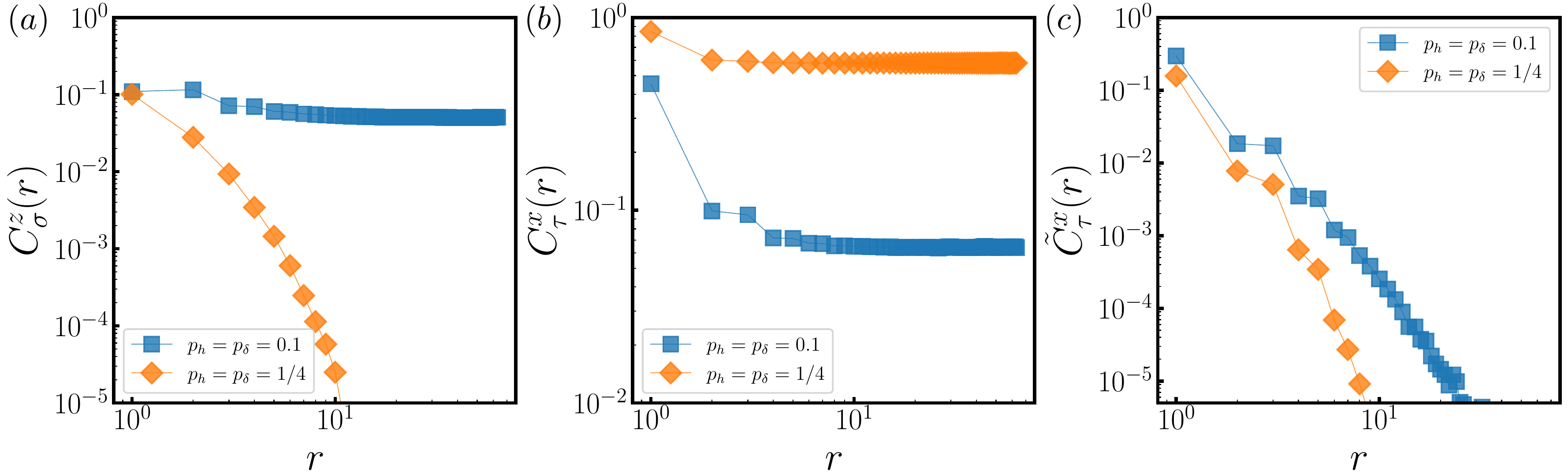}
    \caption{Spin-spin correlations (a) $C_{\sigma}^{z}(r)$ and (b) $C_{\tau}^{x}(r)$ versus the site distance $r$ for two representative points in the steady-state phase diagram shown in Fig.~\ref{fig:sm_z4v2_diagram}(a). (c) The connected correlation $\tilde{C}_{\tau}^{x}(r)$ for the same parameters. Simulations are performed with a system of size $L = 128$ under open boundary conditions.}
    \label{fig:sm_z4v2_correlation}
\end{figure}

For simplicity, we first consider the line $p_{\delta} = 0$. 
As displayed in Fig.~\ref{fig:sm_z4v2_main}(b1), the half-chain entanglement entropy $S_\text{Half}$ grows logarithmically with the system size $L$ at $p_{h} = 0$ ($p_{h} = 1/5$) characterized by an effective central charge $c_\text{eff} = 0.192(2)$ [$c_\text{eff} = 0.092(2)$] while $S_\text{Half}$ saturates to a constant when $L$ is large at other $p_{h}$ values. 
This suggests a percolation transition between two area-law phases near $p_{h} = 1/5$. 
To determine the critical point, we employ the tripartite mutual information $I_{3}$
\begin{equation}
    I_{3} = S_{A}+S_{B}+S_{C}-S_{AB}-S_{BC}-S_{AC}+S_{ABC} \,,
\end{equation}
where the whole system is equally partitioned into another four subsystems, $A|B|C|D$ [see Fig.~\ref{fig:sm_z4v2_main}(a1)]. 
In Fig.~\ref{fig:sm_z4v2_main}(a1), we calculate $I_{3}$ for various $L$ as a function of $p_{h}$.
The perfect data collapse of $I_{3}$ using $\nu = 4/3$ further confirms the percolation transition at the critical point $p_{h} = 1/5$. 
Interestingly, as illustrated in Fig.~\ref{fig:sm_z4v2_main}(c1), a small occurring probability $p_{h}$ causes the string operator $O_{\tau}(r)$ changes immediately from algebraic to exponential decay implying that the single-site perturbation $\tau_{2i-1}^{x}$ gaps out the $\tau$ degrees of freedom and leads to an area-law phase. 
Furthermore, the residue entanglement entropy in the purification dynamics retains $S(t) = 1$ at late time for $p_{h} < 1/5$. 
It can be explained by the $\mathbb{Z}_{2}$ SSB in $\sigma$ degrees of freedom [note that $C_{\sigma}^{z}(r) \sim \text{const}$ at $p_{h} = p_{\delta} = 0.1$, see Fig.~\ref{fig:sm_z4v2_correlation}]. 
When $p_{h} > 1/5$, the symmetry breaking in $\sigma$ degrees of freedom is restored [note that $C_{\sigma}^{z}(r)$ decays exponentially at $p_{h} = p_{\delta} = 0.4$, see Fig.~\ref{fig:sm_z4v2_correlation}] and the corresponding phase is trivially gapped both in $\tau$ and $\sigma$ degrees of freedom as evidenced by a vanishing residue entropy $S(t \gg L) = 0$ [see blue lines in Fig.~\ref{fig:sm_z4v2_main}(d1)]. 

At last, the same simulations are performed for the case of $p_{h} = p_{\delta}$ and similar observations are made in Fig.~\ref{fig:sm_z4v2_main}(a2-d2). 
We also notice that the percolation line $5p_{h} + p_{\delta} = 1$ can be obtained by setting $p_{h} = p_{t}$ as the transition is just the SSB-PM transition happened in the $\sigma$ chain.

\end{document}